\documentclass[nofootinbib,twocolumn,pra,superscriptaddress]{revtex4-2}
\usepackage{graphicx} %
\usepackage{hyperref,xcolor}
\usepackage{verbatim}
\usepackage{amsmath}
\usepackage{amsthm}
\usepackage{amsfonts}
\usepackage{braket}
\usepackage{bbm}
\usepackage{bm}
\hypersetup{colorlinks=true, linkcolor=blue!80!black, urlcolor=blue!80!black, citecolor=blue!80!black}
\usepackage{subcaption}
\usepackage[nameinlink]{cleveref}

\usepackage{tikz,xcolor}
\definecolor{darkgreen}{rgb}{0.0, 0.5, 0.0}
\usetikzlibrary{shapes.geometric, arrows, positioning}

\tikzstyle{arrow} = [thick,->,>=stealth,fill=darkgreen, line width=2mm] %

\newtheorem*{maxcut}{MaxCut}
\newtheorem*{po}{Portfolio Optimization}

\begin{document}
\title{End-to-End Protocol for High-Quality QAOA Parameters with Few Shots} %

\author{Tianyi Hao}
\altaffiliation{These authors contributed equally to this work. Correspondence should be addressed to \texttt{zichang.he@jpmchase.com}}
\affiliation{Global Technology Applied Research, JPMorganChase, New York, NY 10017, USA}

\author{Zichang He}
\altaffiliation{These authors contributed equally to this work. Correspondence should be addressed to \texttt{zichang.he@jpmchase.com}}
\affiliation{Global Technology Applied Research, JPMorganChase, New York, NY 10017, USA}

\author{Ruslan Shaydulin}
\email{ruslan.shaydulin@jpmchase.com}
\affiliation{Global Technology Applied Research, JPMorganChase, New York, NY 10017, USA}
\author{Jeffrey Larson} %
\affiliation{Mathematics and Computer Science Division, Argonne National Laboratory, Lemont, IL 60439, USA}
\author{Marco Pistoia}
\affiliation{Global Technology Applied Research, JPMorganChase, New York, NY 10017, USA}

\begin{abstract}
    The quantum approximate optimization algorithm (QAOA) is a quantum heuristic for combinatorial optimization that has been demonstrated to scale better than state-of-the-art classical solvers for some problems. For a given problem instance, QAOA performance depends crucially on the choice of the parameters. 
    While average-case optimal parameters are available in many cases, meaningful performance gains can be obtained by fine-tuning these parameters for a given instance. 
    This task is especially challenging, however, when the number of circuit executions (shots) is limited. 
    In this work, we develop an end-to-end protocol that combines multiple parameter settings and fine-tuning techniques.
    We use large-scale numerical experiments to optimize the protocol for the shot-limited setting and observe that optimizers with the simplest internal model (linear) perform best. 
    We implement the optimized pipeline on a trapped-ion processor using up to $32$ qubits and $5$ QAOA layers, and we demonstrate that the pipeline is robust to small amounts of hardware noise. 
    To the best of our knowledge, these are the largest demonstrations of QAOA  
    parameter fine-tuning on a trapped-ion processor in terms of 2-qubit gate count.

\end{abstract}

\maketitle

\section{Introduction}

\begin{figure*}[t]
    \includegraphics{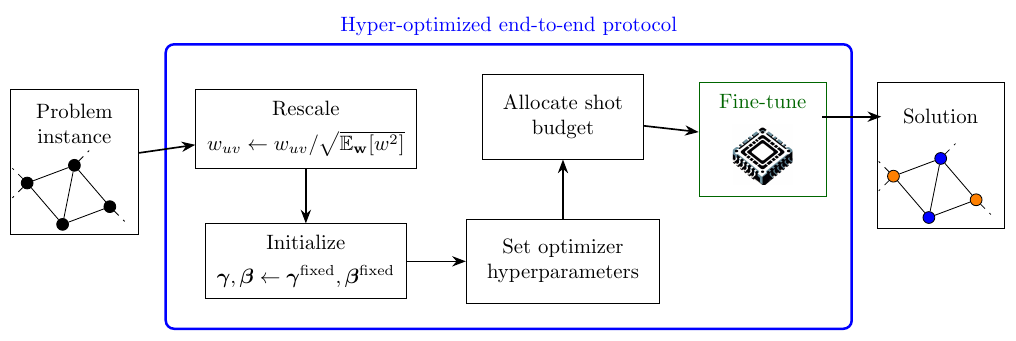}
  \caption{\textbf{Overview of our protocol.} Given a problem instance, we first follow~\cite{sureshbabuParameterSettingQuantum2024} to rescale the weights and then set parameters to known good initial points~\cite{basso2022qaoaskmaxcut, wurtz2021fixed}, based on the parameter concentration property of QAOA~\cite{akshayParameterConcentrationsQuantum2021, shaydulinParameterTransferQuantum2023}. We then set the hyperparameters of optimizers and allocate the shot budget. Both hyperparameter choice and the shot budget allocation are informed by extensive numerical investigation detailed in this paper.
  } 
  \label{fig:protocol}
\end{figure*}

Quantum computing has shown great promise in tackling computational problems that are difficult for classical computers. 
Among such problems, combinatorial optimization problems are of particular interest due to their ubiquity in fields including finance, logistics, and operations research and due to the existence of quantum algorithms offering speedups~\cite{quant-ph/9607014,montanaro2018quantum,montanaro2020quantum,Somma2008,Wocjan2008}. 
The quantum approximate optimization algorithm (QAOA)~\cite{hogg2000quantum,farhi2014quantum,Hadfield2019} 
is a prominent quantum heuristic that has been demonstrated to achieve better scaling than state-of-the-art classical solvers for certain combinatorial optimization problems, including maximum 8-satisfiability~\cite{boulebnane2022solving} and low autocorrelation binary sequence~\cite{shaydulin2023evidence} problems. QAOA solves an optimization problem by preparing a parameterized quantum state such that upon measuring it, a high-quality solution is obtained with high probability.
To apply QAOA to a given problem, QAOA parameters must be set.

The performance of QAOA is highly sensitive to the choice of these parameters, and its parameter optimization has been widely studied in the community~\cite{he2024_dro,hao2024variational}. For many problem classes, optimal parameters have been derived in the infinite-size limit and empirically demonstrated to achieve good performance for finite-sized instances~\cite{basso2022qaoaskmaxcut,boulebnane2022solving,2110.10685,wurtz2021fixed,sureshbabuParameterSettingQuantum2024}. Even when rigorous theoretical analysis is out of reach, one fixed set of empirically obtained QAOA parameters can work well for most instances~\cite{shaydulin2023evidence}. 
Nonetheless, 
there is still nontrivial variation in the optimal parameters between instances, which is often amplified by adding weights to the problem.
Thus, fine-tuning the average-case or infinite-size parameters for a given instance is necessary to fully exploit the algorithm's potential~\cite{lubinski2023optimization,shaydulinParameterTransferQuantum2023,Shaydulin2019}.
Fine-tuning these parameters is challenging, however, especially when the number of circuit executions (shots) is limited, as is often the case with current quantum hardware.

Shots are the fundamental currency of near-term quantum computation. One ``shot'' represents an execution of a quantum circuit followed by a measurement. Each optimization iteration in QAOA requires hundreds or thousands of shots to minimize the sampling error of an expectation value evaluation~\cite{kahani2023novel}. 
The limitations of near-term quantum devices, such as their scarcity, frequent and time-consuming recalibration, and slow operation time, constrain quantum resource availability and, thus, the total number of shots available for an algorithm run. This constraint is particularly pronounced in atomic platforms such as trapped-ion and neutral atom quantum processors, where measurement time is on the same order of magnitude as gate time~\cite{Moses2023RaceTrack,Bluvstein2023,2406.02501}.

The optimization of QAOA parameters presents a significant challenge if the number of shots is limited. This challenge is exacerbated by the fundamental limits that quantum mechanics imposes on the cost of computing gradients of quantum circuits~\cite{2305.13362}. The high cost of computing the gradient motivates the use of derivative-free optimization (DFO), which typically either assumes a deterministic objective~\cite{LMW2019} or requires a high number of shots to converge~\cite{Larson2016,stormoriginal,Shashaani2018,scriva2024challenges}. As a consequence of these challenges and despite the recent progress~\cite{sachdeva2024quantum,menickelly2023latency,ito2023latency,gu2021adaptive,arrasmith2020operator,Sung2020,zhu2023optimizing,moussa2023resource,kubler2020adaptive,arrasmith2020operator,polloreno2022qaoa,cheng2024quantum}, the problem of optimizing parameterized quantum circuits with a small number of shots remains open.

In this work, we propose and implement an end-to-end protocol for obtaining high-quality QAOA parameters with a small number of shots. Our protocol integrates multiple techniques to reduce the cost of parameter optimization, as shown in \Cref{fig:protocol}. These techniques include previously studied ones such as initialization with instance-independent or ``fixed'' parameters~\cite{basso2022qaoaskmaxcut,2110.10685,boulebnane2022solving,shaydulinParameterTransferQuantum2023,sureshbabuParameterSettingQuantum2024} and rescaling of weighted problems~\cite{shaydulinParameterTransferQuantum2023,sureshbabuParameterSettingQuantum2024}, as well as new components to carry out parameter fine-tuning in the shot-frugal setting, including optimizer selection, hyperparameter tuning, and shot budget allocation.
We use extensive numerical experiments to optimize all aspects of our pipeline. In doing so, we demonstrate that the optimizer with the simplest internal model
is the best option in shot-frugal scenarios. 
Our protocol performs well without further classical configuring when given a new problem instance.

We demonstrate the effectiveness of our protocol by deploying it on trapped-ion quantum devices applied to QAOA circuits with up to $32$ qubits. Our protocol is optimized in noiseless simulation but is robust to small amounts of noise and achieves good performance on hardware. For example, in one instance of $20$-qubit 3-regular MaxCut with five QAOA layers, the parameter setting protocol achieves up to 56.61\% relative 
approximation ratio
(AR) improvement in noiseless simulation while it holds 46.88\% relative AR
improvement under hardware noise. We observe that as the circuit size grows, the noise becomes too strong, and the performance of the protocol deteriorates.

\section{Background}

The quantum approximate optimization algorithm (QAOA)~\cite{hogg2000quantum,farhi2014quantum,Hadfield2019} solves combinatorial optimization problems by preparing a parameterized quantum circuit such that upon measuring it, high-quality solutions are obtained with high probability.
The circuit
is defined by two operators, problem Hamiltonian $\bm{H}_P$ and mixer Hamiltonian $\bm{H}_M$, a hyperparameter $p$, and an initial state $\ket{\bm{\psi}_0}$:
\begin{align}
    \left\vert {{{\bm{\psi }}}}({{{\bm{\gamma }}}},{{{\bm{\beta }}}})\right\rangle ={e}^{-i{\beta }_{p}{{{{\bm{H}}}}}_{M}}{e}^{-i{\gamma }_{p}{{{{\bm{H}}}}}_{P}}\ldots {e}^{-i{\beta }_{1}{{{{\bm{H}}}}}_{M}}{e}^{-i{\gamma }_{1}{{{{\bm{H}}}}}_{P}}\left\vert {{{{\bm{\psi }}}}}_{0}\right\rangle ,
\end{align}
where $\bm{\gamma}=[\gamma_0,\cdots,\gamma_p]$ and $\bm{\beta}=[\beta_0,\cdots,\beta_p]$ are free parameters. 
As the number of layers $p$ approaches infinity, the QAOA circuit with appropriate parameters $\ket{\bm{\psi}(\bm{\gamma}, \bm{\beta})}$ approaches the ground state of the problem Hamiltonian $\bm{H}_P$, and the corresponding energy approaches the optimal value of the problem's objective function.
The parameters $\bm{\gamma}, \bm{\beta}$ are typically obtained by using a classical optimizer that iteratively updates them based on the measurement outcomes:
\begin{equation}
    \min_{\bm{\gamma}, \bm{\beta}} \braket{\bm{\psi}(\bm{\gamma}, \bm{\beta})|\bm{H}_P|\bm{\psi}(\bm{\gamma}, \bm{\beta})},
\end{equation}
where $\braket{\bm{\psi}(\bm{\gamma}, \bm{\beta}) | \bm{H}_P | \bm{\psi}(\bm{\gamma}, \bm{\beta})}$ is the expectation value of the energy.

The optimization of variational parameters within the QAOA framework presents a significant challenge. On the one hand, the objective function in practical implementations is inherently stochastic. In each QAOA iteration, the expectation value $\braket{\bm{\psi}(\bm{\gamma}, \bm{\beta}) | \bm{H}_P | \bm{\psi}(\bm{\gamma}, \bm{\beta})}$ is estimated and given to the optimizer as the objective by sampling numerous measurement results from the QAOA state:
\begin{equation}\label{eq:energy_obj}
\begin{split}
    \braket{\bm{\psi}(\bm{\gamma}, \bm{\beta})|\bm{H}_P|\bm{\psi}(\bm{\gamma}, \bm{\beta})} & \approx \frac{1}{M} \sum_{i}^M f(\bm{x}_i), \text{with} \\
    \bm{x}_i & \sim {\lvert \bm{\psi}(\bm{\gamma},\bm{\beta}) \rvert }^2,
\end{split}
\end{equation}
where $f(\bm{x})$ is the asscoiated problem value of a measurement $\bm{x}$.

Each measurement represents an execution of the entire circuit, and the number of circuit executions used to estimate a state is referred to as the number of shots. The most interesting and promising use cases involve limited shots of the stochastic objective. 
Based on the central limit theorem~\cite{billingsley2017probability}, 
with $M$ shots, the standard deviation of estimated energy becomes $\frac{\sigma_0}{\sqrt{M}}$, where
\begin{equation}
    \sigma_0 = \left(\braket{\bm{\psi}(\bm{\gamma}, \bm{\beta})|\bm{H}^2_P|\bm{\psi}(\bm{\gamma}, \bm{\beta})} - 
    {\braket{\bm{\psi}(\bm{\gamma}, \bm{\beta})|\bm{H}_P|\bm{\psi}(\bm{\gamma}, \bm{\beta})}}^2\right)^\frac{1}{2}. 
\end{equation}
This uncertainty in the estimated energy caused by finite sampling becomes significant when the number of shots is small, such as below 1,000, which renders the parameter optimization challenging.

On the other hand, there is limited access to gradients of the QAOA objective, which means practical QAOA experiments have to rely on optimization techniques that do not utilize derivative information, a task that is naturally more complex than gradient-based optimization~\cite{LMW2019}. 
In the absence of gradients, quantum computing researchers have turned to derivative-free optimization (DFO) techniques as the classical optimization approaches in their QAOA work. DFO approaches can coarsely be categorized into direct-search and model-based methods. Direct-search methods evaluate the objective at a geometric pattern of points around a candidate point. If a better point is observed, the best point is updated; otherwise, the displacement is decreased in the geometric pattern. Model-based methods also evaluate points near a candidate point and use these evaluations to build various local or global models of the objective being optimized.

Applying DFO methods that are designed for deterministic objectives to stochastic objectives often leads to suboptimal performance. Rigorous convergence guarantees for stochastic DFO methods typically demand a substantial number of samples to accurately construct or adjust optimization models~\cite{Larson2016,stormoriginal,Shashaani2018}; applying them to quantum optimization tasks will likely be difficult. There is ongoing research aimed at adapting deterministic DFO methods to better accommodate the inherent noise within stochastic objectives, striving for a balance between robustness and sample efficiency~\cite{Shi2024}.

In this paper, we apply QAOA to weighted maximum cut (MaxCut) and portfolio optimization (PO) problems. We now briefly discuss how QAOA is instantiated to be applied to these problems. 

\begin{maxcut}
Given an undirected graph $G=(V, E)$ with an edge weight $w_{uv}$ associated with each edge $(u,v)\in E$, 
find $\bm{s} \in \{-1,1\}^{|V|}$, that will 
\begin{align*}
    \text{\upshape maximize } f(\bm{s}) = \sum_{(u,v)\in E}\frac{w_{uv}}{2} (1-s_us_v).
\end{align*}
\end{maxcut}

Mapping spin variables $s_i$ onto the spectrum of Pauli $\bm{Z}$ matrices, we obtain the Hamiltonian that encodes the MaxCut problem on qubits:
\begin{align}
    \bm{H}_P = \sum_{(u,v)\in E}\frac{w_{uv}}{2} (\bm{I} - \bm{Z}_{u}\bm{Z}_{v}).
\end{align}

We use the Pauli $\bm{X}$ mixer Hamiltonian when applying QAOA to the MaxCut problem:
\begin{align}
    \bm{H}_M = \sum_{i} \bm{X}_i\label{eq:x_mixer}.
\end{align}

The initial state $\ket{\bm{\psi}_0}$ is set to be the ground state of the mixer Hamiltonian, which for $\bm{H}_M$ in \Cref{eq:x_mixer} is
\begin{align}
    \ket{\bm{\psi}_0} = \ket{+}^{\otimes n}.
\end{align}

\begin{po}
Given assets with expected returns $\bm{\mu} \in \mathbb{R}^n$
and covariance 
$\bm{\Sigma} \in \mathbb{R}^{n\times n}$, a risk factor $q \in \mathbb{R}$, and a budget $K \in \mathbb{N}$, find $\bm{x} \in \{0,1\}^n$ that will
\begin{align*}
    &\text{\upshape minimize } && f(\bm{x}) = q \bm{x}^T \bm{\Sigma} \bm{x} - \bm{\mu}^T \bm{x}\\
    &\text{\upshape subject to } && \bm{1}^T \bm{x} = K.
\end{align*}
\end{po}

Mapping binary variables $x_i$ to the Pauli $\bm{Z}$ matrices as $x_i \rightarrow (\bm{I} - \bm{Z}_i)/2$, we get the Hamiltonian
\begin{align}
    {{{{\bm{H}}}}}_{P}=\frac{1}{2}q\mathop{\sum}\limits_{i < j}{W}_{ij}{{{{\bm{Z}}}}}_{i}{{{{\bm{Z}}}}}_{j}-\frac{1}{2}\mathop{\sum}\limits_{i}\left(q\mathop{\sum}\limits_{j}{\Sigma}_{ij}-{\mu }_{i}\right){{{{\bm{Z}}}}}_{i}+c
\end{align}
encoding PO on qubits, where $c=\frac{1}{2}{\sum }_{i}(q{\sum }_{j = i}{W}_{ij}-{\mu }_{i})$ is a constant. To preserve the Hamming weight of the state, we use an $XY$ mixer with a 1-dimensional ring connectivity defined as
\begin{equation}
    \bm{H}_M = \sum_i \sum_{j=i+1} \bm{X}_{i}\bm{X}_{j} + \bm{Y}_{i}\bm{Y}_{j}.
\end{equation}
The initial state is prepared as a Dicke state~\cite{bartschi2019deterministic}, which is a superposition of all feasible (i.e., Hamming weight $K$) bitstrings with an equal probability.

Given a solution $\bm{s}$ or $\bm{x}$ to the problem, we use approximation ratio ($\mathrm{AR}$) to quantify the quality of the solution. For MaxCut, it is defined as
\begin{equation}
  \mathrm{AR}(\bm{s}) = \frac{f(\bm{s})-f_{\rm min}}{f_{\rm max}-f_{\rm min}},
\end{equation}
where $f_{\rm min}$ and $f_{\rm max}$ are the minimum and maximum  value of $f(s)$, respectively; that is,
\begin{equation}    
\begin{aligned}
    f_{\rm min} = \min_{\bm{s}} f(\bm{s}), \\
    f_{\rm max} = \max_{\bm{s}} f(\bm{s}).
\end{aligned}
\end{equation}
For PO, we need to take constraints into consideration:
\begin{equation}\label{eq:ar_po}
  \mathrm{AR}(\bm{x}) = 
  \begin{cases}
  \frac{f(\bm{x})-f_{\rm max}}{f_{\rm min}-f_{\rm max}}, & \sum_i x_i = K, \\
  0, & \sum_i x_i \neq K,
  \end{cases}
\end{equation}
where $f_{\rm min}$ and $f_{\rm max}$ are 
\begin{equation}
\begin{aligned}
    f_{\rm min} = \min_{\sum_i {x_i} = K} f(\bm{x}), \\
    f_{\rm max} = \max_{\sum_i {x_i} = K} f(\bm{x}).
\end{aligned}
\end{equation}

We also use the metric of relative AR improvement, defined as 
\begin{align}\label{eq:relative_ar_improve}
    \frac{\mathrm{AR}(x) - \mathrm{AR}_{ini}}{\mathrm{AR}_{opt} - \mathrm{AR}_{ini}},
\end{align}
where $\mathrm{AR}_{ini}$ and $\mathrm{AR}_{opt}$ are the approximations ratios corresponding to the collection of solutions produced by QAOA circuits with initial parameters $\bm{\gamma}_{ini}, \bm{\beta}_{ini}$ and optimal parameters $\bm{\gamma}_{opt}, \bm{\beta}_{opt}$:
\begin{equation}
\begin{aligned}
    \mathrm{AR}_{ini} = \mathbb{E}_{x\sim\ket{\bm{\psi}(\bm{\gamma}_{ini},\bm{\beta}_{ini})}} [\mathrm{AR}(x)]\\
    \mathrm{AR}_{opt} = \mathbb{E}_{x\sim\ket{\bm{\psi}(\bm{\gamma}_{opt},\bm{\beta}_{opt})}} [\mathrm{AR}(x)].
\end{aligned}
\end{equation}
From a practical perspective, the optimal parameters refer to the set of parameters we can empirically find that lead to the best objective function value.
In our evaluations, we obtain the optimal parameters by performing noiseless optimizations with unlimited shots.

\section{Results}
We now present our results. We begin by summarizing our protocol in \Cref{sec:protocol_summary} and briefly introducing existing techniques our protocol uses. We proceed by describing the rest of its components, with optimizer selection in \Cref{sec:optimizer}, hyperparameter study in \Cref{sec:hyperparameter}, and budget allocation in \Cref{sec:budget}. We then present the performance of the protocol on trapped-ion hardware in \Cref{sec:hardware}.

\subsection{End-to-end protocol for QAOA parameter optimization}\label{sec:protocol_summary}

\Cref{fig:protocol} shows an overview of our protocol. Given a problem instance, we first follow~\cite{sureshbabuParameterSettingQuantum2024} to rescale the weights. We divide the objective function 
\begin{equation}
    \sqrt{\frac{1}{|E_2|}\sum_{i,j}w_{ij}^2 +\frac{1}{|E_1|}\sum_{i}w_{i}^2 },
\end{equation}
where $|E_2|$ is the number of quadratic terms in the objective function and $|E_1|$ is the number of first-order terms. This rescaling rule can also be extended for problems with higher-degree terms. 

The difficulty of optimizing the parameters in QAOA heavily depends on the initial point selection. It has been shown for several problem settings that
the optimized parameters for different problem instances are approximately equal~\cite{basso2022qaoaskmaxcut,shaydulinParameterTransferQuantum2023,sureshbabuParameterSettingQuantum2024,shaydulin2023evidence,boulebnane2022solving,2110.10685}.
Consequently, the averaged optimized parameters from several problem instances serve as a high-quality initial point. Thus, we use the parameters given in~\cite{wurtz2021fixed} for unweighted MaxCut with 3-regular graphs as our initial points for MaxCut and follow the empirical observation in Ref.~\cite{sureshbabuParameterSettingQuantum2024} to 
use the averaged optimized parameters for the Sherrington-Kirkpatrick model~\cite{basso2022qaoaskmaxcut} for PO. 
The values of $\bm{\gamma}$ and $\bm{\beta}$ are further rescaled so that they are on the same scale for the optimizer.

The rest of the components concern optimizer selection, hyperparameter tuning, and shot budget allocation, which we describe in detail in the following subsections.

\subsection{Optimizer choice}\label{sec:optimizer}

\begin{figure}
    \centering
    \includegraphics{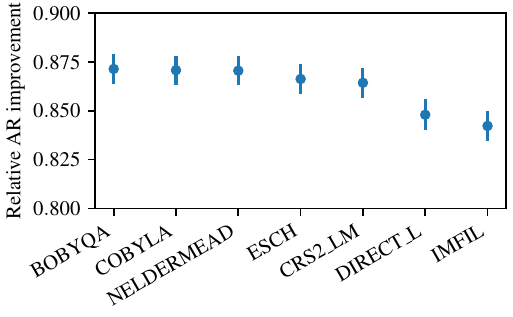}
    \caption{Performance comparison of common derivative-free optimizers optimizing $p$=1 QAOA circuits for 60 random PO instances. Each optimizer is evaluated under different budget allocation strategies with a total budget of 10,000 shots, and model-based optimizers also have their initial step size grid searched. The best-performed hyperparameter combination is then used to plot this figure. The metric is the mean relative AR improvement with standard error over instances. The optimizers are arranged in descending order of mean AR. BYBOQA, COBYLA, and NELDER\_MEAD achieve very comparable performance. The mean relative AR improvement optimized by SPSA is 0.385, which is too low to be included in the figure.}
    \label{fig:optimizer}
\end{figure}

We conduct an evaluation of various optimization methods under the shot-frugal setting.
Specifically, we examine the performance of 
COBYLA~\cite{COBYLA},
BOBYQA~\cite{BOBYQA}, 
NELDER\_MEAD~\cite{NELDERMEAD}, 
ESCH~\cite{ESCH},
DIRECT\_L~\cite{DIRECT_L},
CRS2\_LM~\cite{CRS2_LM},
SPSA~\cite{SPSA},
GSLS~\cite{GSLS},
and
IMFIL~\cite{IMFIL}. These methods have been considered in the quantum
optimization context and are available in optimization packages
NLopt~\cite{NLopt}, PyBOBYQA~\cite{cartis2019improving}, SciPy~\cite{2020SciPy-NMeth}, PDFO~\cite{PDFO}, and Scikit-Quant~\cite{lavrijsen2020classical}. 
Some other methods we also tested but either work very similarly to one of the above methods or perform undesirably include
UOBYQA~\cite{UOBYQA}, 
NEWUOA~\cite{NEWUOA},
LINCOA~\cite{Powell2015}, 
and 
SNOBFIT~\cite{SNOBFIT}.
We compute the energy landscapes of 60 random $p$=1 PO instances and efficiently test the optimization quality of
each method. %

\begin{figure*}[t]
    \centering
    \begin{subfigure}{0.49\textwidth}
        \centering
        \includegraphics[width=\linewidth]{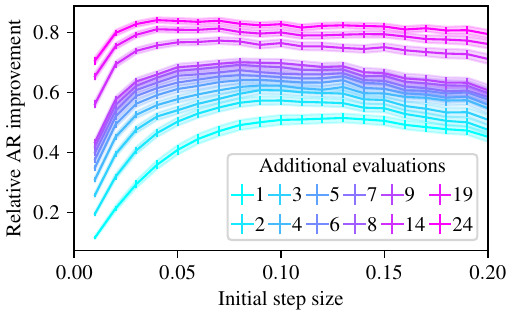}
        \subcaption{MaxCut}\label{fig:maxcut-rhobeg}
    \end{subfigure}
    \begin{subfigure}{0.49\textwidth}
        \centering
        \includegraphics[width=\linewidth]{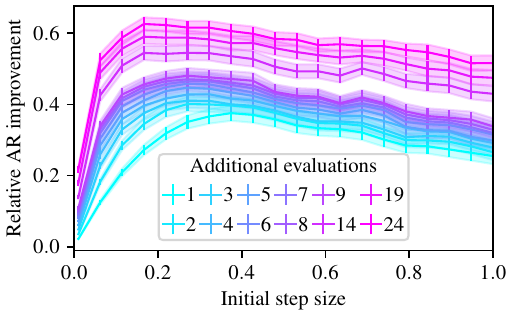}
        \subcaption{PO}\label{fig:po-rhobeg}
    \end{subfigure}
    \caption{Mean relative AR improvement (with standard error over instances) of COBYLA on $p=5$ QAOA instances as a function of initial step size, assuming an infinite shot budget. The label of each line represents the number of function evaluations allowed after the initial evaluations.  We observe that MaxCut, having a more accurate initialization strategy, requires a smaller initial step size than does PO.
    With a given problem and initialization strategy, COBYLA is generally not sensitive to the initial step size.}
    \label{fig:rhobeg}
\end{figure*}

Some of the tested methods have dozens of hyperparameters that can be adjusted
before running; other methods have only one or two. A complete study of the
performance of each method as its hyperparameters change is far beyond the scope
of this manuscript. Instead, we focus our numerical studies on the allocation of the total budget, an additional hyperparameter they all share that is crucial to our restricted setting. This is particularly important if the optimization method needs a few initial function evaluations before being able to make a prediction since more function evaluations lead to fewer shots per evaluation. In addition, the ``initial step size'' of model-based methods, which determines the spread of the initial pattern of points, is a crucial hyperparameter that impacts the performance significantly. We also vary it and choose the best-performing one for these methods in our benchmarking.
We test a total of 1,460 optimization configurations, and each configuration is evaluated 5 times with different sampling seeds on each of the instances. Please refer to \Cref{sec:oscar} for the details of the efficient performance evaluation.

\Cref{fig:optimizer} shows the performance comparison of the tested methods using mean relative AR improvement (\cref{eq:relative_ar_improve}) as the metric.
We see that BOBYQA, COBYLA, and NELDER\_MEAD  perform the best in this setting among all tested methods. We choose COBYLA over NELDER\_MEAD since the former is considered an improved version of the latter~\cite{COBYLA}. Between BOBYQA and COBYLA, we choose the latter, attributing to the fact that a simple model minimizes the number of initial function evaluations and maximizes the number of shots per evaluation. COBYLA assumes a linear model and needs only $2p+1$ initial function evaluations to build the model for $2p$ parameters in a $p$-layer QAOA. BOBYQA follows almost the same strategy as COBYLA except that it assumes a quadratic model, which requires up to $\frac{1}{2}(2p+1)(2p+2)$ initial function evaluations to fully determine the model. We adopt the default setting of $4p+1$ initial evaluations, which is still almost twice as many as COBYLA requires in the large $p$ limit.
In the small $p$ regions, the performance of BOBYQA and COBYLA are comparable (see more details in \Cref{sec:appendix}).
However, the number of shots per evaluation for BOBYQA approaches half of that for COBYLA as we increase the number of QAOA layers ($p$). The substantial uncertainty due to shotted evaluations outweighs the benefits of fitting a more refined model. Similarly, other methods generally have more complicated assumptions and are thus more demanding in terms of the number of evaluations. The improved accuracy in the predictions does not compensate for the loss resulting from the significant decrease in the number of shots per evaluation.

\subsection{Hyperparameter selection}\label{sec:hyperparameter}

We now fix COBYLA as our optimization method and investigate its hyperparameter choice. COBYLA only has one hyperparameter to tune besides the budget allocation strategy, which is the initial step size (``rhobeg''). 
As previously mentioned, this hyperparameter determines the distance between the initial point and other initial evaluations, which is crucial for establishing an accurate linear model while using few evaluations. 
We run exact simulations of 100 random 12-qubit MaxCut instances and 60 random 12-qubit PO instances for $p \in \{1,\ldots,5\}$ and plot the optimization results under varying initial step sizes. 

In \Cref{fig:rhobeg}, we plot the mean relative AR improvement with standard error over instances as a function of initial step size, where each curve represents the performance at the given number of function evaluations after the initial evaluations. The performance converges toward the maximum possible relative improvement as we increase the function evaluations, as expected. 
We observe that MaxCut, having a more accurate initialization strategy, requires a smaller initial step size than does PO.
With a given problem and initialization strategy, we notice that COBYLA is generally not sensitive to the initial step size, even when the number of evaluations is very low. The same trend can be observed across different numbers of QAOA layers ($p$), which we show in \Cref{fig:more_rhobeg} in \Cref{sec:appendix}.
For subsequent experiments, we use an initial step size of 0.1 for MaxCut and 0.5 for PO.

\subsection{Budget allocation}\label{sec:budget}

We now study the budget allocation strategy. We primarily focus on a per-instance total budget of $10,000$ shots. This value of total budget was chosen to match the constraints of trapped-ion hardware. Prior QAOA experiments on the device used in this work used 2000, 5200, and 7800 shots and $p\leq 2$~\cite{decross2023qubit, Moses2023RaceTrack}.
For comparison, previous hardware demonstration on a superconducting processor used 25,000 shots per evaluation and 6 evaluations per optimizer iteration~\cite{harrigan2021quantum}. The number of shots per evaluation is inversely proportional to the number of total function evaluations:
\begin{equation}
    \#\text{shots/evaluation} = \left\lfloor \frac{\text{shot budget}=10,000}{\#\text{evaluations allowed}} \right\rfloor.\label{eq:shots-per-eval}
\end{equation}
We perform hyperparameter grid searches on the number of shots per function evaluation versus the maximum number of function evaluations on 1,000 random 12-qubit MaxCut instances for $p \in \{2,3,4,5\}$.

\begin{figure}[t]
    \centering
    \includegraphics[width=\linewidth]{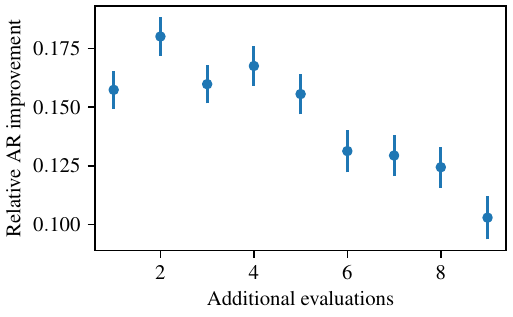}
    \caption{Mean relative AR improvement (with standard error over instances) of optimizing 1,000 $p=5\ n=12$ MaxCut instances as a function of the number of additional evaluations after the first $2p+1$ initial evaluations. The number of shots per evaluation is divided evenly from a total budget of 10,000 shots. Empirically, we observe that 2 iterations after the initial evaluations work the best on the average case.
    }
    \label{fig:budget-slice}
\end{figure}

\Cref{fig:budget-slice} shows the mean relative AR improvement with standard error over instances as a function of the maximum number of function evaluations given to the optimizer, starting from the required number of initial function evaluations for building the linear model in COBYLA. We observe that in exchange for more function evaluations, the rapidly reduced number of shots per evaluation significantly impacts the measurement accuracy and optimizer behavior. With a good initial point, the optimizer can get close to the optimal point in a few iterations after the initial evaluations. Thus, the best strategy is to maximize the number of shots per evaluation so that the optimizer can rely on the initial evaluations to accurately predict candidate points. Empirically, we observe that 2 iterations after the initial evaluations work the best on the average case.

\begin{figure}[t]
    \centering
    \includegraphics[width=\linewidth]{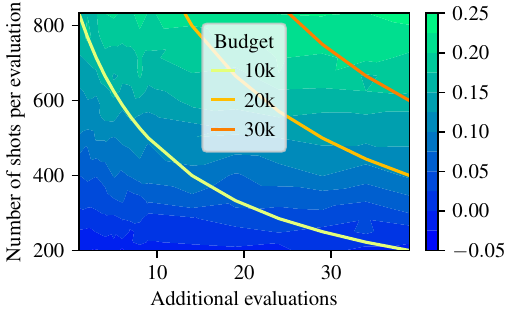}
    \caption{Contour plot of optimizing 1,000 $p=5\ n=12$ MaxCut instances spanned by the number of additional evaluations and the number of shots per evaluation. The color represents mean relative AR improvement, and the three lines correspond to a total budget of 10k, 20k, and 30k, respectively.
    }
    \label{fig:budget-heatmap}
\end{figure}

\Cref{fig:budget-heatmap} shows a contour plot spanned by the number of additional evaluations and the number of shots per evaluation. The color represents mean relative AR improvement, and the three lines correspond to a total budget of 10k, 20k, and 30k, respectively. Focusing on the 10k budget line, we make the same observation as in \Cref{fig:budget-slice}: smaller number of steps and higher per-evaluation shot budget
give the best results. Following the contours, We see that increasing the number of evaluations has little return, in contrast to the steady gain in increasing the number of shots per evaluation. If the number of shots is as low as 200, the optimizer will even find lower-quality parameters than the initial ones due to the extreme variance in sampling the objective function values. We expect the budget allocation strategy to be the same for similar budgets, whereas a substantially higher budget can allow more evaluations. 
We show additional results for $p \in \{2, 3, 4\}$ in \Cref{fig:more_budget_slice} and \Cref{fig:more_budget_heatmap} in \Cref{sec:appendix}.
We can also derive the minimum shot budget requirement from these contour figures, where the budget results in approximately zero improvement in the relative AR.

\subsection{Hardware demonstrations}\label{sec:hardware}
\begin{figure}[t]
    \centering
    \includegraphics[width = \linewidth]{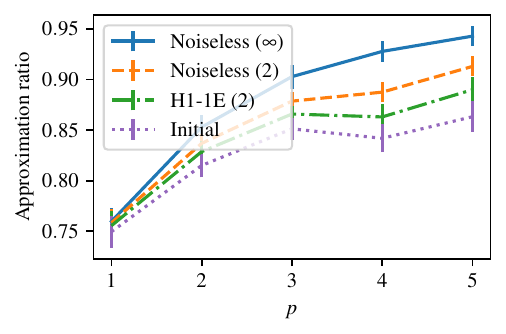}
    \caption{Optimizing QAOA parameters with $p \in \{1,\ldots,5\}$ for $n=12$ weighted graphs. Error bars represent the standard error of the mean AR over 5 instances. 
    The Noiseless $(\infty)$ line optimizes the QAOA parameter under the noiseless backend and with an unlimited number of iterations. In other words, it represents the performance of the best achievable parameters. The Noiseless $(2)$ line optimizes the QAOA parameter under the noiseless backend and with two additional iterations, representing the results of the protocol in noiseless simulation. The H1-1E $(2)$ line optimizes with two additional iterations under an emulator backend
    , which mimics the H1-1 quantum device, representing the deployment of the protocol in a practical scenario. The Initial line presents the performance under the fixed initial parameter.
    } 
    \label{fig:hardware_maxcut_N12}
\end{figure}
\begin{figure*}[t]
    \centering
    \includegraphics[width = \linewidth]{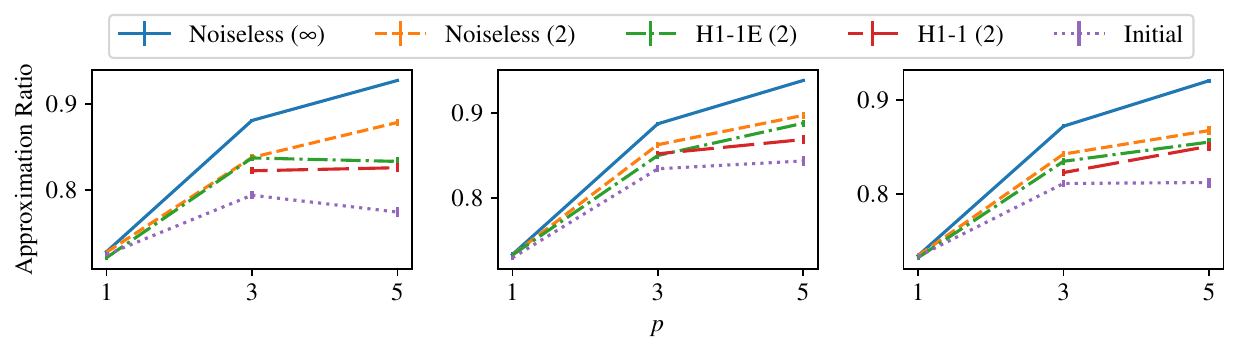}
    \caption{Optimizing QAOA parameters with $p \in \{1,3,5\}$ for three $n=20$ MaxCut instances. Error bars represent the standard error of the mean AR over limited shots. An error bar is estimated as $\frac{\sigma}{\sqrt{M}}$, where $\sigma$ is the standard deviation of the approximation ratio associated with the exact QAOA state vector and $M$ is the number of shots. The added H1-1 (2) line optimizes for 2 additional iterations after the first $2p+1$ initial evaluations. For the description of other labels, please refer to the caption of \Cref{fig:hardware_maxcut_N12}.} 
    \label{fig:hardware_maxcut_N20}
\end{figure*}
\begin{figure}[t]
    \centering
    \includegraphics[width = \linewidth]{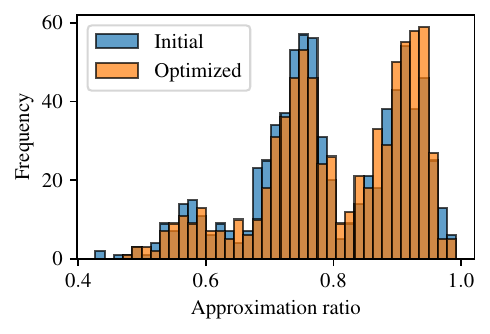}
    \caption{Shot frequency of solving the $n=32$ MaxCut with $p=5$ QAOA in H2-1. The initial bins represent $769$ samples from the $p = 5$ QAOA state with initial parameters. The optimized bins represent $769$ samples with fine-tuned parameters.} 
    \label{fig:hardware_maxcut_N32}
\end{figure}
\begin{figure}[t]
    \centering
    \includegraphics[width = \linewidth]{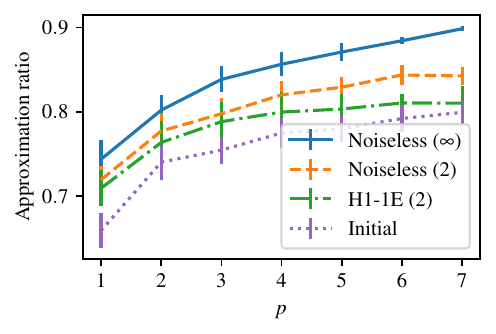}
    \caption{Optimizing QAOA parameters with $p \in \{1,\ldots,7\}$ for $n=10$ PO instances. Error bars represent the standard error of the mean AR over $5$ instances. For a description of labels, please refer to the caption of \Cref{fig:hardware_maxcut_N12}.} 
    \label{fig:hardware_po}
\end{figure}

We demonstrate the effectiveness of our parameter setting protocol with Quantinuum's noisy emulator of H1-1 (denoted as H1-1E), as well as the H1-1~\cite{H1} and H2-1~\cite{H2} quantum processors. Each optimization has a total budget of 10k shots across all evaluations for one problem instance.

We first validate the protocol on MaxCut problems. In \Cref{fig:hardware_maxcut_N12}, we report the average performance over five $n=12$ graphs with QAOA depth up to $5$. The Noiseless $(\infty)$ line optimizes the QAOA parameter under the noiseless backend and with an unlimited number of iterations. In other words, it represents the performance of the best achievable parameters. The Noiseless $(2)$ line optimizes the QAOA parameter under the noiseless backend and with two additional iterations after the first $2p+1$ initial evaluations, representing the results of the protocol in noiseless simulation. The H1-1E $(2)$ line optimizes with two additional iterations with an emulator backend, which mimics the H1-1 quantum device, representing the deployment of the protocol in a practical scenario. The Initial line presents the performance under the fixed initial parameter. The results validate the effectiveness of the protocol, showing that under different $p$, the protocol can improve the parameter quality nontrivially. At $p=1$, all the fine-tuning results are close to the initial since the initial parameters are of very high quality. The gap between Noiseless $(2)$ and H1-1E $(2)$ quantifies the impact of the hardware noise. 

In \Cref{fig:hardware_maxcut_N20}, we utilize all the qubits in the H1-1 processor and validate the protocol on three $n=20$ instances. The hardware results are shown on the H1-1 line. The proposed protocol performs well on quantum hardware, confirming the emulator results. In these selected instances, the initial ARs of $p=5$ are close to $p=3$ because the quality of initial parameters is relatively poor for these instances, highlighting the significance of the instance-level fine-tuning. The proposed protocol achieves up to 56.61\% relative AR improvement in noiseless simulation while obtaining 46.88\% relative AR improvement under hardware noise.

We use our protocol to optimize one $n=32$ MaxCut instance on the H2-1 processor. The ARs with the standard error over shots are shown in \Cref{tab:hardware_maxcut_N32}. The ``Hardware evaluation'' column shows the results directly reported by the hardware
optimization. The “Exact evaluation” column shows the results evaluated with a noiseless state vector simulator using the parameters found by the hardware optimization. 
Since the circuit size is larger than in the previous experiments, the gap between the hardware and noiseless simulation also becomes larger. Nonetheless, for $p=5$, $n=32$ and a QAOA circuit with $240$ two-qubit gates, the protocol is still able to optimize the parameter beyond the high-quality initial parameters. We plot the observed distributions with initial and optimized parameters in \Cref{fig:hardware_maxcut_N32}.
\begin{table}[t]
    \centering
    \begin{tabular}{|c|c|c|}
        \hline
        Parameter     & Hardware evaluation & Exact evaluation \\
        \hline
        Initial       & 0.7963 (0.1367) & 0.8084 \\
        H2-1 + 2 iter & 0.8137 (0.1335) & 0.8197 \\
        Noiseless + 2 iter & N/A & 0.8654 \\
        \hline
    \end{tabular}
    \caption{Numerical simulation and hardware demonstration results of $p=5$ QAOA on an $n=32$ MaxCut instance. Values in parentheses are the standard error of the mean AR over limited shots. ``2 iter'' denotes the results, including 2 iterations after the first $2p+1$ initial evaluations. The ``Hardware evaluation'' column shows the results directly reported by the hardware optimization. The ``Exact evaluation'' column shows the results evaluated with a noiseless state vector simulator using the parameters found by the hardware optimization.}
    \label{tab:hardware_maxcut_N32}
\end{table}

We also use PO to verify our findings and numerically demonstrate the protocol's general applicability. PO is a harder problem setting because the initial parameter quality is generally poorer than MaxCut's. In addition, the constraint-preserving mixer has a larger overhead in the circuit.
We validate the protocol with H1-1E for five $n=10$ PO instances with $p$ up to $7$, as shown in \Cref{fig:hardware_po}. The QAOA circuit has $97$ two-qubit gates for initial state preparation and $65$ two-qubit gates for one QAOA layer. We consistently observe the improved AR over the initial parameter setting. Meanwhile, as $p$ becomes larger, the circuit encounters
an increasingly greater amount of noise, and the AR performance gap between the noisy and noiseless simulation becomes larger. 
For the $p=7$ experiments with $552$ two-qubit gates, the relative AR improvement in the noisy simulation is $11.02\%$ while it is $43.52\%$ in the noiseless simulation. 
Parameter optimization in a highly noisy environment is challenging even under the fine-tuning setup, suggesting the necessity of deploying error suppression strategies and developing error correction techniques in larger-scale hardware experiments.

\section{Discussion}
\begin{figure*}[t]
    \centering
    \begin{subfigure}{0.49\textwidth}
        \centering
        \includegraphics[width=\linewidth]{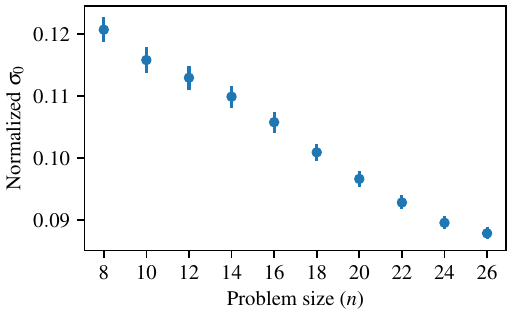}
        \subcaption{}\label{fig:std_over_N}
    \end{subfigure}
    \begin{subfigure}{0.49\textwidth}
        \centering
        \includegraphics[width=\linewidth]{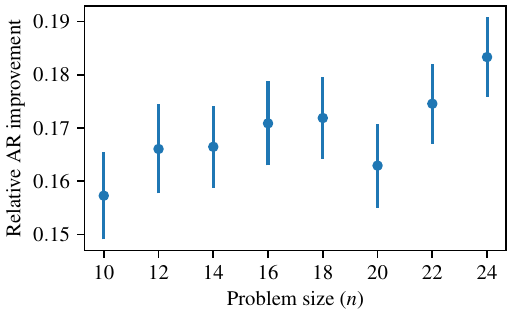}
        \subcaption{}\label{fig:ar_over_N}
    \end{subfigure}
    \caption{(a) Normalized standard deviation of the estimated energy (single-shot) and (b) relative AR improvement of our protocol (10,000 shots) as a function of problem size $n$ averaged over 1,000 $p=5$ MaxCut instances. Error bars denote the standard error. Considering the absolute change is not substantial, we believe the dip at $n=20$ in (b) is just a normal deviation.}
\end{figure*}

QAOA has shown algorithmic speedup over the classical state of the art for some problems. High-quality parameters are necessary to realize its quantum speedup. However, although optimal parameters in the average cases are usually available, instance-level fine-tuning is necessary for maximizing the QAOA performance. In this work, we confirm that such fine-tunings can reliably improve the QAOA performance beyond the high-quality initialization with optimization steps linear in the QAOA layer $p$.

We propose an end-to-end protocol for QAOA parameter setting. Focusing on weighted combinatorial optimization problems within the same family, we assume we have a fixed parameter initialization schedule for the unweighted problems from the same family. We first rescale the weighted problem instance such that the parameters for unweighted problems can be applicable. 
Then, we need to tune the hyperparameters of the classical optimizer. Under the shot-limited cases, we benchmark 12 optimization algorithms and found that COBYLA and BOBYQA perform the best. Considering a realistic shot budget of 10k per problem instance, we use the simplest method, COBYLA. To tune the hyperparameters of COBYLA, we observe that a fixed initial step size works stably in QAOA with different layers for problems of different sizes. Then, we determine the maximum number of iterations of COBYLA to optimize the budget allocation. Under the setting of 10k shots, we assume an equally allocated budget and perform a grid search on the number of shots for one sample. We set the number of steps to be 2 and observe that it consistently works well.

Our protocol and shot budget assumption work for different problem sizes. \Cref{fig:std_over_N} shows the normalized standard deviation of the estimated energy induced by finite sampling for MaxCut of varying problem sizes. Each data point is averaged over 1000 instances. We observe that as the problem size increases, the standard deviation decreases, indicating that fewer shots are needed to maintain the same level of sampling error. We further show the relative AR improvement of our protocol as a function of the problem size in \Cref{fig:ar_over_N}, which confirms that our protocol stays effective at larger problem sizes. 

Our protocol can be applied to different problems. We test the protocol with MaxCut and PO instances. For MaxCut, we solve a 32-qubit problem with up to 5 layers of QAOA and perform hardware demonstrations with both H2-1 and H1-1 devices. For PO, we solve 10-qubit problems with up to 7 layers of QAOA and demonstrate in the emulator of the H1-1 device. The protocol is effective and robust to noise. For example, in one instance of 20-qubit MaxCut with $p=5$, the fine-tuning strategy achieves up to 56.61\% relative AR improvement in noiseless simulation while it holds 46.88\% relative AR improvement under hardware noise.

This protocol provides an end-to-end solution for QAOA parameter settings on a quantum device. 
The protocol is also generally applicable when we have many more shots, 
allowing for more fine-tuning iterations. In such scenarios, several open and interesting directions remain. For example, we can dynamically allocate the shots, allowing each iteration to have a different number of shots. Additionally, we can assign the shot budget to problem instances that perform relatively poorly under the fixed initial parameters. Finally, we remark that our techniques can be combined with other qubit-reuse compilation techniques (e.g., Ref.~\cite{decross2023qubit}) to dramatically increase the size of the problem that can be tackled.

\section{Methods}
We now present our approach to test our protocol through both numerical simulations and quantum demonstrations. We detail our methods for noiseless and noisy simulations, hardware experiments, and the specific combinatorial optimization problems we considered. 

\subsection{Evaluation and data generation}\label{sec:evaluation}
We perform both numerical simulations and quantum demonstrations to derive, tune, and validate our protocol. For noiseless simulations, we employ an optimized MPI-enabled QAOA simulation package, QOKit~\cite{Lykov2023}, to simulate up to 32 qubits on 2 Nvidia A100-80GB GPUs. For noisy simulations, we use the Quantinuum H1 emulator~\cite{H1}.
For hardware experiments, we obtain the results on the Quantinuum H1-1 and H2-1 quantum processors. H1-1 and H2-1 are trapped-ion devices that have full connectivity and high gate fidelity. For detailed specifications of H1-1 and H2-1, please refer to~\cite{H1} and~\cite{H2}.
Whenever the experiment setting is not exact (by exact, we mean noiseless with unlimited shots), we first obtain optimized parameters under the experiment settings and then report the results by exactly evaluating these parameters. This measure is to present the quality of the optimized parameters truthfully and fairly.

We consider two combinatorial optimization problems: weighted maximum cut (MaxCut) and portfolio optimization (PO). For MaxCut, we generate random 3-regular graphs with random edge weights sampled from a mixture of three Gaussian distributions $\{(\mathcal{N}(0, 1), 0.5), (\mathcal{N}(5, 2), 0.3), (\mathcal{N}(10, 1), 0.2)\}$, where 0.5, 0.3, and 0.2 are the probabilities of sampling from the corresponding Gaussian distribution. 
For PO, we use historical stock market data from Yahoo! Finance as the source for generating the expected return vectors and the covariance matrices. The 60 instances we use for each $n$ are from the first 28 days of each month from January 2015 to December 2019 with $n$ selected stocks from S\&P 500 companies that have complete data during that period.

For the optimizer performance comparison, we assume a total budget of 10,000 shots, vary the number of function evaluations from $2p+2$ to 20 and allocate the shot budget evenly to each evaluation. We also grid search the initial step size for supported optimizers from 0.05 to 0.5 with an increment of 0.05. Further details are described in the next subsection. For initial step size studies, we assume an infinite shot budget. We grid search the initial step size from 0.01 to 0.2 with an increment of 0.01 for MaxCut and from 0.01 to 1 with an increment of approximately 0.05 for PO. For budget allocation studies, we set the number of function evaluations from $2p+2$ to 50 with an increment of 1 between $2p+2$ and 20 and an increment of 5 between 20 and 50. The number of shots per evaluation is set using \Cref{eq:shots-per-eval}.

\subsection{Low-cost optimizer benchmarking}\label{sec:oscar}
\begin{figure*}[t]
    \centering
    \begin{subfigure}{0.246\textwidth}
        \centering
        \includegraphics[width=0.96\textwidth, trim={0.73cm, 0, 0.1cm, 0}, clip]{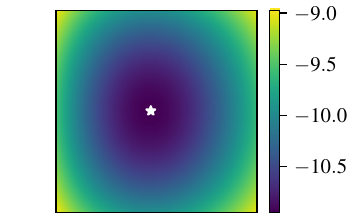}
        \subcaption{}\label{fig:landscape(a)}
    \end{subfigure}
    \begin{subfigure}{0.246\textwidth}
        \centering
        \includegraphics[width=0.96\textwidth, trim={0.73cm, 0, 0.1cm, 0}, clip]{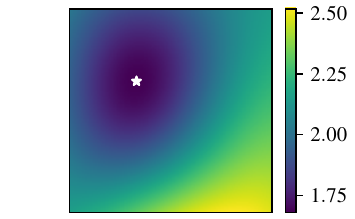}
        \subcaption{}\label{fig:landscape(b)}
    \end{subfigure}
    \begin{subfigure}{0.246\textwidth}
        \centering
        \includegraphics[width=\textwidth, trim={0.73cm, 0.15cm, 0.1cm, 0.05cm}, clip]{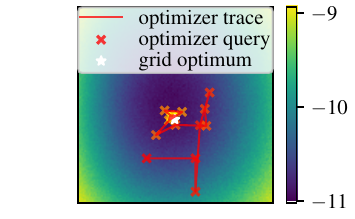}
        \subcaption{}\label{fig:landscape(c)}
    \end{subfigure}
    \begin{subfigure}{0.246\textwidth}
        \centering
        \includegraphics[width=0.96\textwidth, trim={0.73cm, 0, 0.1cm, 0}, clip]{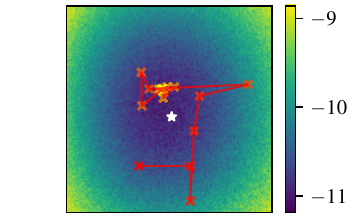}
        \subcaption{}\label{fig:landscape(d)}
    \end{subfigure}
    \caption{(a) Energy mean landscape of a $p=1$ PO instance with a resolution (points along each dimension) of 128 by 128 over the $\frac{\pi}{4}$ by $\frac{\pi}{4}$ region where the center is the initial point we use in our experiments. (b) Energy standard deviation landscape of the same instance. (c) Sampled landscape showing the effect of 5,000 shots per evaluation. An optimization trajectory by COBYLA is overlaid for demonstration. The red to yellow marks represent COBYLA's queries (function evaluations), where the lower left point is the start and the yellow mark is the end. They are connected by a red line (``optimizer trace'') to indicate the order. (d) Sampled landscape with 500 shots per evaluation. We see that finite sampling noise has a serious impact on the optimization quality.
    }
    \label{fig:landscape}
\end{figure*}

We employ the quantum optimization helper package OSCAR~\cite{OSCAR, hao2024variational} to conduct a computationally tractable evaluation of the various optimizers under different budget allocation strategies. For each QAOA instance, OSCAR computes a discrete energy landscape and then interpolates discrete points on the landscape to answer subsequent function evaluations without actually simulating the circuit. This enables us to quickly evaluate numerous optimization configurations with manageable costs.

Nonetheless, the discrete landscape computation is a grid search, which means its complexity is exponential in the number of points along each dimension. Although OSCAR supports reconstructing a landscape with a small number of sampled points, we choose to evaluate each point on the landscape with exact simulation to avoid approximations. 
Thus, this approach is viable only for low-depth QAOA landscapes. For our optimizer benchmark experiments, we use $p=1$ instances and a resolution of 128 by 128 over the $\frac{\pi}{4}$ by $\frac{\pi}{4}$ region where the center is the initial point we use in our experiments.
\Cref{fig:landscape(a)} shows an example energy landscape we use. Notice that the local optimum (white star in the figure) is very close to the initial point (center of the figure), showing the effectiveness of our initialization strategy.

To simulate the energy with finite sampling errors, in addition to the energy landscapes, we compute the standard deviation landscapes to realize fast energy evaluation with an arbitrary shot. 

\Cref{fig:landscape(b)} shows an example standard deviation landscape we use. \Cref{fig:landscape(c)} and \Cref{fig:landscape(d)} show two landscapes sampled from the energy and the standard deviation landscapes with 5,000 and 500 shots per evaluation, respectively. We see that the finite sampling adds a noticeable salt-like noise to the landscapes, and fewer shots lead to heavier noise. To demonstrate the impact on optimization, we overlay COBYLA's evaluations on top of the landscape. Note that for demonstration purposes, we start from a lower left point instead of the center, which is the initial point we use in our experiments. We observe that the optimized point (yellow mark in the figure) is seriously affected by the sampling noise.

\section*{Data availability}

The data for reproducing figures used in this paper is available at \cite{datalink}.

\section*{Code availability}

The code for reproducing the data and figures used in this paper is available at \cite{codelink}.

\section*{Acknowledgments}
This material is based upon work supported in part by the U.S.~Department of Energy, Office of Science, under contract number DE-AC02-06CH11357
and the Office of Science, Office of Advanced Scientific Computing Research, Accelerated Research for Quantum Computing program.

The authors thank Danylo Lykov for his help with numerical experiments. TH, ZH, RS, and MP  thank their colleagues at Global Technology Applied Research of JPMorganChase for their support and helpful discussions.

\newpage
\bibliography{reference.bib}

\begin{thebibliography}{70}%
\makeatletter
\providecommand \@ifxundefined [1]{%
 \@ifx{#1\undefined}
}%
\providecommand \@ifnum [1]{%
 \ifnum #1\expandafter \@firstoftwo
 \else \expandafter \@secondoftwo
 \fi
}%
\providecommand \@ifx [1]{%
 \ifx #1\expandafter \@firstoftwo
 \else \expandafter \@secondoftwo
 \fi
}%
\providecommand \natexlab [1]{#1}%
\providecommand \enquote  [1]{``#1''}%
\providecommand \bibnamefont  [1]{#1}%
\providecommand \bibfnamefont [1]{#1}%
\providecommand \citenamefont [1]{#1}%
\providecommand \href@noop [0]{\@secondoftwo}%
\providecommand \href [0]{\begingroup \@sanitize@url \@href}%
\providecommand \@href[1]{\@@startlink{#1}\@@href}%
\providecommand \@@href[1]{\endgroup#1\@@endlink}%
\providecommand \@sanitize@url [0]{\catcode `\\12\catcode `\$12\catcode
  `\&12\catcode `\#12\catcode `\^12\catcode `\_12\catcode `\%12\relax}%
\providecommand \@@startlink[1]{}%
\providecommand \@@endlink[0]{}%
\providecommand \url  [0]{\begingroup\@sanitize@url \@url }%
\providecommand \@url [1]{\endgroup\@href {#1}{\urlprefix }}%
\providecommand \urlprefix  [0]{URL }%
\providecommand \Eprint [0]{\href }%
\providecommand \doibase [0]{https://doi.org/}%
\providecommand \selectlanguage [0]{\@gobble}%
\providecommand \bibinfo  [0]{\@secondoftwo}%
\providecommand \bibfield  [0]{\@secondoftwo}%
\providecommand \translation [1]{[#1]}%
\providecommand \BibitemOpen [0]{}%
\providecommand \bibitemStop [0]{}%
\providecommand \bibitemNoStop [0]{.\EOS\space}%
\providecommand \EOS [0]{\spacefactor3000\relax}%
\providecommand \BibitemShut  [1]{\csname bibitem#1\endcsname}%
\let\auto@bib@innerbib\@empty
\bibitem [{\citenamefont {Sureshbabu}\ \emph {et~al.}(2024)\citenamefont
  {Sureshbabu}, \citenamefont {Herman}, \citenamefont {Shaydulin},
  \citenamefont {Basso}, \citenamefont {Chakrabarti}, \citenamefont {Sun},\
  and\ \citenamefont {Pistoia}}]{sureshbabuParameterSettingQuantum2024}%
  \BibitemOpen
  \bibfield  {author} {\bibinfo {author} {\bibfnamefont {S.~H.}\ \bibnamefont
  {Sureshbabu}}, \bibinfo {author} {\bibfnamefont {D.}~\bibnamefont {Herman}},
  \bibinfo {author} {\bibfnamefont {R.}~\bibnamefont {Shaydulin}}, \bibinfo
  {author} {\bibfnamefont {J.}~\bibnamefont {Basso}}, \bibinfo {author}
  {\bibfnamefont {S.}~\bibnamefont {Chakrabarti}}, \bibinfo {author}
  {\bibfnamefont {Y.}~\bibnamefont {Sun}},\ and\ \bibinfo {author}
  {\bibfnamefont {M.}~\bibnamefont {Pistoia}},\ }\bibfield  {title} {\bibinfo
  {title} {Parameter setting in quantum approximate optimization of weighted
  problems},\ }\href {https://doi.org/10.22331/q-2024-01-18-1231} {\bibfield
  {journal} {\bibinfo  {journal} {Quantum}\ }\textbf {\bibinfo {volume} {8}},\
  \bibinfo {pages} {1231} (\bibinfo {year} {2024})}\BibitemShut {NoStop}%
\bibitem [{\citenamefont {Basso}\ \emph {et~al.}(2022)\citenamefont {Basso},
  \citenamefont {Farhi}, \citenamefont {Marwaha}, \citenamefont {Villalonga},\
  and\ \citenamefont {Zhou}}]{basso2022qaoaskmaxcut}%
  \BibitemOpen
  \bibfield  {author} {\bibinfo {author} {\bibfnamefont {J.}~\bibnamefont
  {Basso}}, \bibinfo {author} {\bibfnamefont {E.}~\bibnamefont {Farhi}},
  \bibinfo {author} {\bibfnamefont {K.}~\bibnamefont {Marwaha}}, \bibinfo
  {author} {\bibfnamefont {B.}~\bibnamefont {Villalonga}},\ and\ \bibinfo
  {author} {\bibfnamefont {L.}~\bibnamefont {Zhou}},\ }\bibfield  {title}
  {\bibinfo {title} {{The Quantum Approximate Optimization Algorithm at High
  Depth for MaxCut on Large-Girth Regular Graphs and the
  Sherrington-Kirkpatrick Model}},\ }in\ \href
  {https://doi.org/10.4230/LIPIcs.TQC.2022.7} {\emph {\bibinfo {booktitle}
  {17th Conference on the Theory of Quantum Computation, Communication and
  Cryptography (TQC 2022)}}},\ \bibinfo {series} {Leibniz International
  Proceedings in Informatics (LIPIcs)}, Vol.\ \bibinfo {volume} {232},\
  \bibinfo {editor} {edited by\ \bibinfo {editor} {\bibfnamefont
  {F.}~\bibnamefont {Le~Gall}}\ and\ \bibinfo {editor} {\bibfnamefont
  {T.}~\bibnamefont {Morimae}}}\ (\bibinfo  {publisher} {Schloss Dagstuhl --
  Leibniz-Zentrum f{\"u}r Informatik},\ \bibinfo {address} {Dagstuhl,
  Germany},\ \bibinfo {year} {2022})\ pp.\ \bibinfo {pages}
  {7:1--7:21}\BibitemShut {NoStop}%
\bibitem [{\citenamefont {Wurtz}\ and\ \citenamefont
  {Lykov}(2021)}]{wurtz2021fixed}%
  \BibitemOpen
  \bibfield  {author} {\bibinfo {author} {\bibfnamefont {J.}~\bibnamefont
  {Wurtz}}\ and\ \bibinfo {author} {\bibfnamefont {D.}~\bibnamefont {Lykov}},\
  }\bibfield  {title} {\bibinfo {title} {The fixed angle conjecture for {QAOA}
  on regular {MaxCut} graphs},\ }\href
  {https://doi.org/10.48550/arXiv.2107.00677} {\bibfield  {journal} {\bibinfo
  {journal} {arXiv:2107.00677}\ } (\bibinfo {year} {2021})}\BibitemShut
  {NoStop}%
\bibitem [{\citenamefont {Akshay}\ \emph {et~al.}(2021)\citenamefont {Akshay},
  \citenamefont {Rabinovich}, \citenamefont {Campos},\ and\ \citenamefont
  {Biamonte}}]{akshayParameterConcentrationsQuantum2021}%
  \BibitemOpen
  \bibfield  {author} {\bibinfo {author} {\bibfnamefont {V.}~\bibnamefont
  {Akshay}}, \bibinfo {author} {\bibfnamefont {D.}~\bibnamefont {Rabinovich}},
  \bibinfo {author} {\bibfnamefont {E.}~\bibnamefont {Campos}},\ and\ \bibinfo
  {author} {\bibfnamefont {J.}~\bibnamefont {Biamonte}},\ }\bibfield  {title}
  {\bibinfo {title} {Parameter concentrations in quantum approximate
  optimization},\ }\href {https://doi.org/10.1103/PhysRevA.104.L010401}
  {\bibfield  {journal} {\bibinfo  {journal} {Physical Review A}\ }\textbf
  {\bibinfo {volume} {104}},\ \bibinfo {pages} {L010401} (\bibinfo {year}
  {2021})}\BibitemShut {NoStop}%
\bibitem [{\citenamefont {Shaydulin}\ \emph {et~al.}(2023)\citenamefont
  {Shaydulin}, \citenamefont {Lotshaw}, \citenamefont {Larson}, \citenamefont
  {Ostrowski},\ and\ \citenamefont
  {Humble}}]{shaydulinParameterTransferQuantum2023}%
  \BibitemOpen
  \bibfield  {author} {\bibinfo {author} {\bibfnamefont {R.}~\bibnamefont
  {Shaydulin}}, \bibinfo {author} {\bibfnamefont {P.~C.}\ \bibnamefont
  {Lotshaw}}, \bibinfo {author} {\bibfnamefont {J.}~\bibnamefont {Larson}},
  \bibinfo {author} {\bibfnamefont {J.}~\bibnamefont {Ostrowski}},\ and\
  \bibinfo {author} {\bibfnamefont {T.~S.}\ \bibnamefont {Humble}},\ }\bibfield
   {title} {\bibinfo {title} {Parameter transfer for quantum approximate
  optimization of weighted {MaxCut}},\ }\href {https://doi.org/10.1145/3584706}
  {\bibfield  {journal} {\bibinfo  {journal} {ACM Transactions on Quantum
  Computing}\ }\textbf {\bibinfo {volume} {4}},\ \bibinfo {pages} {19:1}
  (\bibinfo {year} {2023})}\BibitemShut {NoStop}%
\bibitem [{\citenamefont {D\"{u}rr}\ and\ \citenamefont
  {H\o{}yer}(1996)}]{quant-ph/9607014}%
  \BibitemOpen
  \bibfield  {author} {\bibinfo {author} {\bibfnamefont {C.}~\bibnamefont
  {D\"{u}rr}}\ and\ \bibinfo {author} {\bibfnamefont {P.}~\bibnamefont
  {H\o{}yer}},\ }\bibfield  {title} {\bibinfo {title} {A quantum algorithm for
  finding the minimum},\ }\href
  {https://doi.org/10.48550/arXiv.quant-ph/9607014} {\bibfield  {journal}
  {\bibinfo  {journal} {arXiv:quant-ph/9607014}\ } (\bibinfo {year}
  {1996})}\BibitemShut {NoStop}%
\bibitem [{\citenamefont {Montanaro}(2018)}]{montanaro2018quantum}%
  \BibitemOpen
  \bibfield  {author} {\bibinfo {author} {\bibfnamefont {A.}~\bibnamefont
  {Montanaro}},\ }\bibfield  {title} {\bibinfo {title} {Quantum-walk speedup of
  backtracking algorithms},\ }\href {https://doi.org/10.4086/toc.2018.v014a015}
  {\bibfield  {journal} {\bibinfo  {journal} {Theory Of Computing}\ }\textbf
  {\bibinfo {volume} {14}},\ \bibinfo {pages} {1} (\bibinfo {year}
  {2018})}\BibitemShut {NoStop}%
\bibitem [{\citenamefont {Montanaro}(2020)}]{montanaro2020quantum}%
  \BibitemOpen
  \bibfield  {author} {\bibinfo {author} {\bibfnamefont {A.}~\bibnamefont
  {Montanaro}},\ }\bibfield  {title} {\bibinfo {title} {Quantum speedup of
  branch-and-bound algorithms},\ }\href
  {https://doi.org/10.1103/PhysRevResearch.2.013056} {\bibfield  {journal}
  {\bibinfo  {journal} {Physical Review Research}\ }\textbf {\bibinfo {volume}
  {2}},\ \bibinfo {pages} {013056} (\bibinfo {year} {2020})}\BibitemShut
  {NoStop}%
\bibitem [{\citenamefont {Somma}\ \emph {et~al.}(2008)\citenamefont {Somma},
  \citenamefont {Boixo}, \citenamefont {Barnum},\ and\ \citenamefont
  {Knill}}]{Somma2008}%
  \BibitemOpen
  \bibfield  {author} {\bibinfo {author} {\bibfnamefont {R.~D.}\ \bibnamefont
  {Somma}}, \bibinfo {author} {\bibfnamefont {S.}~\bibnamefont {Boixo}},
  \bibinfo {author} {\bibfnamefont {H.}~\bibnamefont {Barnum}},\ and\ \bibinfo
  {author} {\bibfnamefont {E.}~\bibnamefont {Knill}},\ }\bibfield  {title}
  {\bibinfo {title} {Quantum simulations of classical annealing processes},\
  }\href {https://doi.org/10.1103/physrevlett.101.130504} {\bibfield  {journal}
  {\bibinfo  {journal} {Physical Review Letters}\ }\textbf {\bibinfo {volume}
  {101}} (\bibinfo {year} {2008})}\BibitemShut {NoStop}%
\bibitem [{\citenamefont {Wocjan}\ and\ \citenamefont
  {Abeyesinghe}(2008)}]{Wocjan2008}%
  \BibitemOpen
  \bibfield  {author} {\bibinfo {author} {\bibfnamefont {P.}~\bibnamefont
  {Wocjan}}\ and\ \bibinfo {author} {\bibfnamefont {A.}~\bibnamefont
  {Abeyesinghe}},\ }\bibfield  {title} {\bibinfo {title} {Speedup via quantum
  sampling},\ }\href {https://doi.org/10.1103/physreva.78.042336} {\bibfield
  {journal} {\bibinfo  {journal} {Physical Review A}\ }\textbf {\bibinfo
  {volume} {78}} (\bibinfo {year} {2008})}\BibitemShut {NoStop}%
\bibitem [{\citenamefont {Hogg}\ and\ \citenamefont
  {Portnov}(2000)}]{hogg2000quantum}%
  \BibitemOpen
  \bibfield  {author} {\bibinfo {author} {\bibfnamefont {T.}~\bibnamefont
  {Hogg}}\ and\ \bibinfo {author} {\bibfnamefont {D.}~\bibnamefont {Portnov}},\
  }\bibfield  {title} {\bibinfo {title} {Quantum optimization},\ }\href
  {https://doi.org/10.1016/S0020-0255(00)00052-9} {\bibfield  {journal}
  {\bibinfo  {journal} {Information Sciences}\ }\textbf {\bibinfo {volume}
  {128}},\ \bibinfo {pages} {181} (\bibinfo {year} {2000})}\BibitemShut
  {NoStop}%
\bibitem [{\citenamefont {Farhi}\ \emph {et~al.}(2014)\citenamefont {Farhi},
  \citenamefont {Goldstone},\ and\ \citenamefont {Gutmann}}]{farhi2014quantum}%
  \BibitemOpen
  \bibfield  {author} {\bibinfo {author} {\bibfnamefont {E.}~\bibnamefont
  {Farhi}}, \bibinfo {author} {\bibfnamefont {J.}~\bibnamefont {Goldstone}},\
  and\ \bibinfo {author} {\bibfnamefont {S.}~\bibnamefont {Gutmann}},\
  }\bibfield  {title} {\bibinfo {title} {A quantum approximate optimization
  algorithm},\ }\href {https://doi.org/10.48550/arXiv.1411.4028} {\bibfield
  {journal} {\bibinfo  {journal} {arXiv:1411.4028}\ } (\bibinfo {year}
  {2014})}\BibitemShut {NoStop}%
\bibitem [{\citenamefont {Hadfield}\ \emph {et~al.}(2019)\citenamefont
  {Hadfield}, \citenamefont {Wang}, \citenamefont {O'Gorman}, \citenamefont
  {Rieffel}, \citenamefont {Venturelli},\ and\ \citenamefont
  {Biswas}}]{Hadfield2019}%
  \BibitemOpen
  \bibfield  {author} {\bibinfo {author} {\bibfnamefont {S.}~\bibnamefont
  {Hadfield}}, \bibinfo {author} {\bibfnamefont {Z.}~\bibnamefont {Wang}},
  \bibinfo {author} {\bibfnamefont {B.}~\bibnamefont {O'Gorman}}, \bibinfo
  {author} {\bibfnamefont {E.~G.}\ \bibnamefont {Rieffel}}, \bibinfo {author}
  {\bibfnamefont {D.}~\bibnamefont {Venturelli}},\ and\ \bibinfo {author}
  {\bibfnamefont {R.}~\bibnamefont {Biswas}},\ }\bibfield  {title} {\bibinfo
  {title} {From the quantum approximate optimization algorithm to a quantum
  alternating operator ansatz},\ }\href {https://doi.org/10.3390/a12020034}
  {\bibfield  {journal} {\bibinfo  {journal} {Algorithms}\ }\textbf {\bibinfo
  {volume} {12}},\ \bibinfo {pages} {34} (\bibinfo {year} {2019})}\BibitemShut
  {NoStop}%
\bibitem [{\citenamefont {Boulebnane}\ and\ \citenamefont
  {Montanaro}(2024)}]{boulebnane2022solving}%
  \BibitemOpen
  \bibfield  {author} {\bibinfo {author} {\bibfnamefont {S.}~\bibnamefont
  {Boulebnane}}\ and\ \bibinfo {author} {\bibfnamefont {A.}~\bibnamefont
  {Montanaro}},\ }\bibfield  {title} {\bibinfo {title} {Solving {Boolean}
  satisfiability problems with the quantum approximate optimization
  algorithm},\ }\href {https://doi.org/10.1103/PRXQuantum.5.030348} {\bibfield
  {journal} {\bibinfo  {journal} {PRX Quantum}\ }\textbf {\bibinfo {volume}
  {5}},\ \bibinfo {pages} {030348} (\bibinfo {year} {2024})}\BibitemShut
  {NoStop}%
\bibitem [{\citenamefont {Shaydulin}\ \emph {et~al.}(2024)\citenamefont
  {Shaydulin}, \citenamefont {Li}, \citenamefont {Chakrabarti}, \citenamefont
  {DeCross}, \citenamefont {Herman}, \citenamefont {Kumar}, \citenamefont
  {Larson}, \citenamefont {Lykov}, \citenamefont {Minssen}, \citenamefont {Sun}
  \emph {et~al.}}]{shaydulin2023evidence}%
  \BibitemOpen
  \bibfield  {author} {\bibinfo {author} {\bibfnamefont {R.}~\bibnamefont
  {Shaydulin}}, \bibinfo {author} {\bibfnamefont {C.}~\bibnamefont {Li}},
  \bibinfo {author} {\bibfnamefont {S.}~\bibnamefont {Chakrabarti}}, \bibinfo
  {author} {\bibfnamefont {M.}~\bibnamefont {DeCross}}, \bibinfo {author}
  {\bibfnamefont {D.}~\bibnamefont {Herman}}, \bibinfo {author} {\bibfnamefont
  {N.}~\bibnamefont {Kumar}}, \bibinfo {author} {\bibfnamefont
  {J.}~\bibnamefont {Larson}}, \bibinfo {author} {\bibfnamefont
  {D.}~\bibnamefont {Lykov}}, \bibinfo {author} {\bibfnamefont
  {P.}~\bibnamefont {Minssen}}, \bibinfo {author} {\bibfnamefont
  {Y.}~\bibnamefont {Sun}}, \emph {et~al.},\ }\bibfield  {title} {\bibinfo
  {title} {Evidence of scaling advantage for the quantum approximate
  optimization algorithm on a classically intractable problem},\ }\href
  {https://doi.org/10.1126/sciadv.adm6761} {\bibfield  {journal} {\bibinfo
  {journal} {Science Advances}\ }\textbf {\bibinfo {volume} {10}} (\bibinfo
  {year} {2024})}\BibitemShut {NoStop}%
\bibitem [{\citenamefont {He}\ \emph {et~al.}(2024)\citenamefont {He},
  \citenamefont {Peng}, \citenamefont {Alexeev},\ and\ \citenamefont
  {Zhang}}]{he2024_dro}%
  \BibitemOpen
  \bibfield  {author} {\bibinfo {author} {\bibfnamefont {Z.}~\bibnamefont
  {He}}, \bibinfo {author} {\bibfnamefont {B.}~\bibnamefont {Peng}}, \bibinfo
  {author} {\bibfnamefont {Y.}~\bibnamefont {Alexeev}},\ and\ \bibinfo {author}
  {\bibfnamefont {Z.}~\bibnamefont {Zhang}},\ }\bibfield  {title} {\bibinfo
  {title} {Distributionally robust variational quantum algorithms with shifted
  noise},\ }\href {https://doi.org/10.1109/TQE.2024.3409309} {\bibfield
  {journal} {\bibinfo  {journal} {IEEE Transactions on Quantum Engineering}\
  }\textbf {\bibinfo {volume} {5}},\ \bibinfo {pages} {1} (\bibinfo {year}
  {2024})}\BibitemShut {NoStop}%
\bibitem [{\citenamefont {Hao}\ \emph {et~al.}(2024)\citenamefont {Hao},
  \citenamefont {He}, \citenamefont {Shaydulin}, \citenamefont {Pistoia},\ and\
  \citenamefont {Tannu}}]{hao2024variational}%
  \BibitemOpen
  \bibfield  {author} {\bibinfo {author} {\bibfnamefont {T.}~\bibnamefont
  {Hao}}, \bibinfo {author} {\bibfnamefont {Z.}~\bibnamefont {He}}, \bibinfo
  {author} {\bibfnamefont {R.}~\bibnamefont {Shaydulin}}, \bibinfo {author}
  {\bibfnamefont {M.}~\bibnamefont {Pistoia}},\ and\ \bibinfo {author}
  {\bibfnamefont {S.}~\bibnamefont {Tannu}},\ }\bibfield  {title} {\bibinfo
  {title} {Variational quantum algorithm landscape reconstruction by low-rank
  tensor completion},\ }in\ \href {https://doi.org/10.1109/QCE60285.2024.00139}
  {\emph {\bibinfo {booktitle} {2024 IEEE International Conference on Quantum
  Computing and Engineering (QCE)}}},\ Vol.~\bibinfo {volume} {1}\ (\bibinfo
  {organization} {IEEE},\ \bibinfo {year} {2024})\ pp.\ \bibinfo {pages}
  {1184--1190}\BibitemShut {NoStop}%
\bibitem [{\citenamefont {Boulebnane}\ and\ \citenamefont
  {Montanaro}(2021)}]{2110.10685}%
  \BibitemOpen
  \bibfield  {author} {\bibinfo {author} {\bibfnamefont {S.}~\bibnamefont
  {Boulebnane}}\ and\ \bibinfo {author} {\bibfnamefont {A.}~\bibnamefont
  {Montanaro}},\ }\bibfield  {title} {\bibinfo {title} {Predicting parameters
  for the quantum approximate optimization algorithm for {MaxCut} from the
  infinite-size limit},\ }\href {https://doi.org/10.48550/arXiv.2110.10685}
  {\bibfield  {journal} {\bibinfo  {journal} {arXiv:2110.10685}\ } (\bibinfo
  {year} {2021})}\BibitemShut {NoStop}%
\bibitem [{\citenamefont {Lubinski}\ \emph {et~al.}(2023)\citenamefont
  {Lubinski}, \citenamefont {Coffrin}, \citenamefont {McGeoch}, \citenamefont
  {Sathe}, \citenamefont {Apanavicius},\ and\ \citenamefont
  {Neira}}]{lubinski2023optimization}%
  \BibitemOpen
  \bibfield  {author} {\bibinfo {author} {\bibfnamefont {T.}~\bibnamefont
  {Lubinski}}, \bibinfo {author} {\bibfnamefont {C.}~\bibnamefont {Coffrin}},
  \bibinfo {author} {\bibfnamefont {C.}~\bibnamefont {McGeoch}}, \bibinfo
  {author} {\bibfnamefont {P.}~\bibnamefont {Sathe}}, \bibinfo {author}
  {\bibfnamefont {J.}~\bibnamefont {Apanavicius}},\ and\ \bibinfo {author}
  {\bibfnamefont {D.~E.~B.}\ \bibnamefont {Neira}},\ }\bibfield  {title}
  {\bibinfo {title} {Optimization applications as quantum performance
  benchmarks},\ }\href {https://doi.org/10.48550/arXiv.2302.02278} {\bibfield
  {journal} {\bibinfo  {journal} {arXiv:2302.02278}\ } (\bibinfo {year}
  {2023})}\BibitemShut {NoStop}%
\bibitem [{\citenamefont {Shaydulin}\ \emph {et~al.}(2019)\citenamefont
  {Shaydulin}, \citenamefont {Safro},\ and\ \citenamefont
  {Larson}}]{Shaydulin2019}%
  \BibitemOpen
  \bibfield  {author} {\bibinfo {author} {\bibfnamefont {R.}~\bibnamefont
  {Shaydulin}}, \bibinfo {author} {\bibfnamefont {I.}~\bibnamefont {Safro}},\
  and\ \bibinfo {author} {\bibfnamefont {J.}~\bibnamefont {Larson}},\
  }\bibfield  {title} {\bibinfo {title} {Multistart methods for quantum
  approximate optimization},\ }in\ \href
  {https://doi.org/10.1109/hpec.2019.8916288} {\emph {\bibinfo {booktitle}
  {High Performance Extreme Computing Conference}}}\ (\bibinfo  {publisher}
  {IEEE},\ \bibinfo {year} {2019})\BibitemShut {NoStop}%
\bibitem [{\citenamefont {Kahani}\ and\ \citenamefont
  {Nobakhti}(2023)}]{kahani2023novel}%
  \BibitemOpen
  \bibfield  {author} {\bibinfo {author} {\bibfnamefont {S.~S.}\ \bibnamefont
  {Kahani}}\ and\ \bibinfo {author} {\bibfnamefont {A.}~\bibnamefont
  {Nobakhti}},\ }\bibfield  {title} {\bibinfo {title} {A novel framework for
  shot number minimization in quantum variational algorithms},\ }\href
  {https://doi.org/10.48550/arXiv.2307.04035} {\bibfield  {journal} {\bibinfo
  {journal} {arXiv:2307.04035}\ } (\bibinfo {year} {2023})}\BibitemShut
  {NoStop}%
\bibitem [{\citenamefont {Moses}\ \emph {et~al.}(2023)\citenamefont {Moses},
  \citenamefont {Baldwin}, \citenamefont {Allman}, \citenamefont {Ancona},
  \citenamefont {Ascarrunz}, \citenamefont {Barnes}, \citenamefont
  {Bartolotta}, \citenamefont {Bjork}, \citenamefont {Blanchard}, \citenamefont
  {Bohn}, \citenamefont {Bohnet}, \citenamefont {Brown}, \citenamefont
  {Burdick}, \citenamefont {Burton}, \citenamefont {Campbell}, \citenamefont
  {Campora}, \citenamefont {Carron}, \citenamefont {Chambers}, \citenamefont
  {Chan}, \citenamefont {Chen}, \citenamefont {Chernoguzov}, \citenamefont
  {Chertkov}, \citenamefont {Colina}, \citenamefont {Curtis}, \citenamefont
  {Daniel}, \citenamefont {DeCross}, \citenamefont {Deen}, \citenamefont
  {Delaney}, \citenamefont {Dreiling}, \citenamefont {Ertsgaard}, \citenamefont
  {Esposito}, \citenamefont {Estey}, \citenamefont {Fabrikant}, \citenamefont
  {Figgatt}, \citenamefont {Foltz}, \citenamefont {Foss-Feig}, \citenamefont
  {Francois}, \citenamefont {Gaebler}, \citenamefont {Gatterman}, \citenamefont
  {Gilbreth}, \citenamefont {Giles}, \citenamefont {Glynn}, \citenamefont
  {Hall}, \citenamefont {Hankin}, \citenamefont {Hansen}, \citenamefont
  {Hayes}, \citenamefont {Higashi}, \citenamefont {Hoffman}, \citenamefont
  {Horning}, \citenamefont {Hout}, \citenamefont {Jacobs}, \citenamefont
  {Johansen}, \citenamefont {Jones}, \citenamefont {Karcz}, \citenamefont
  {Klein}, \citenamefont {Lauria}, \citenamefont {Lee}, \citenamefont {Liefer},
  \citenamefont {Lu}, \citenamefont {Lucchetti}, \citenamefont {Lytle},
  \citenamefont {Malm}, \citenamefont {Matheny}, \citenamefont {Mathewson},
  \citenamefont {Mayer}, \citenamefont {Miller}, \citenamefont {Mills},
  \citenamefont {Neyenhuis}, \citenamefont {Nugent}, \citenamefont {Olson},
  \citenamefont {Parks}, \citenamefont {Price}, \citenamefont {Price},
  \citenamefont {Pugh}, \citenamefont {Ransford}, \citenamefont {Reed},
  \citenamefont {Roman}, \citenamefont {Rowe}, \citenamefont {Ryan-Anderson},
  \citenamefont {Sanders}, \citenamefont {Sedlacek}, \citenamefont {Shevchuk},
  \citenamefont {Siegfried}, \citenamefont {Skripka}, \citenamefont {Spaun},
  \citenamefont {Sprenkle}, \citenamefont {Stutz}, \citenamefont {Swallows},
  \citenamefont {Tobey}, \citenamefont {Tran}, \citenamefont {Tran},
  \citenamefont {Vogt}, \citenamefont {Volin}, \citenamefont {Walker},
  \citenamefont {Zolot},\ and\ \citenamefont {Pino}}]{Moses2023RaceTrack}%
  \BibitemOpen
  \bibfield  {author} {\bibinfo {author} {\bibfnamefont {S.~A.}\ \bibnamefont
  {Moses}}, \bibinfo {author} {\bibfnamefont {C.~H.}\ \bibnamefont {Baldwin}},
  \bibinfo {author} {\bibfnamefont {M.~S.}\ \bibnamefont {Allman}}, \bibinfo
  {author} {\bibfnamefont {R.}~\bibnamefont {Ancona}}, \bibinfo {author}
  {\bibfnamefont {L.}~\bibnamefont {Ascarrunz}}, \bibinfo {author}
  {\bibfnamefont {C.}~\bibnamefont {Barnes}}, \bibinfo {author} {\bibfnamefont
  {J.}~\bibnamefont {Bartolotta}}, \bibinfo {author} {\bibfnamefont
  {B.}~\bibnamefont {Bjork}}, \bibinfo {author} {\bibfnamefont
  {P.}~\bibnamefont {Blanchard}}, \bibinfo {author} {\bibfnamefont
  {M.}~\bibnamefont {Bohn}}, \bibinfo {author} {\bibfnamefont {J.~G.}\
  \bibnamefont {Bohnet}}, \bibinfo {author} {\bibfnamefont {N.~C.}\
  \bibnamefont {Brown}}, \bibinfo {author} {\bibfnamefont {N.~Q.}\ \bibnamefont
  {Burdick}}, \bibinfo {author} {\bibfnamefont {W.~C.}\ \bibnamefont {Burton}},
  \bibinfo {author} {\bibfnamefont {S.~L.}\ \bibnamefont {Campbell}}, \bibinfo
  {author} {\bibfnamefont {J.~P.}\ \bibnamefont {Campora}}, \bibinfo {author}
  {\bibfnamefont {C.}~\bibnamefont {Carron}}, \bibinfo {author} {\bibfnamefont
  {J.}~\bibnamefont {Chambers}}, \bibinfo {author} {\bibfnamefont {J.~W.}\
  \bibnamefont {Chan}}, \bibinfo {author} {\bibfnamefont {Y.~H.}\ \bibnamefont
  {Chen}}, \bibinfo {author} {\bibfnamefont {A.}~\bibnamefont {Chernoguzov}},
  \bibinfo {author} {\bibfnamefont {E.}~\bibnamefont {Chertkov}}, \bibinfo
  {author} {\bibfnamefont {J.}~\bibnamefont {Colina}}, \bibinfo {author}
  {\bibfnamefont {J.~P.}\ \bibnamefont {Curtis}}, \bibinfo {author}
  {\bibfnamefont {R.}~\bibnamefont {Daniel}}, \bibinfo {author} {\bibfnamefont
  {M.}~\bibnamefont {DeCross}}, \bibinfo {author} {\bibfnamefont
  {D.}~\bibnamefont {Deen}}, \bibinfo {author} {\bibfnamefont {C.}~\bibnamefont
  {Delaney}}, \bibinfo {author} {\bibfnamefont {J.~M.}\ \bibnamefont
  {Dreiling}}, \bibinfo {author} {\bibfnamefont {C.~T.}\ \bibnamefont
  {Ertsgaard}}, \bibinfo {author} {\bibfnamefont {J.}~\bibnamefont {Esposito}},
  \bibinfo {author} {\bibfnamefont {B.}~\bibnamefont {Estey}}, \bibinfo
  {author} {\bibfnamefont {M.}~\bibnamefont {Fabrikant}}, \bibinfo {author}
  {\bibfnamefont {C.}~\bibnamefont {Figgatt}}, \bibinfo {author} {\bibfnamefont
  {C.}~\bibnamefont {Foltz}}, \bibinfo {author} {\bibfnamefont
  {M.}~\bibnamefont {Foss-Feig}}, \bibinfo {author} {\bibfnamefont
  {D.}~\bibnamefont {Francois}}, \bibinfo {author} {\bibfnamefont {J.~P.}\
  \bibnamefont {Gaebler}}, \bibinfo {author} {\bibfnamefont {T.~M.}\
  \bibnamefont {Gatterman}}, \bibinfo {author} {\bibfnamefont {C.~N.}\
  \bibnamefont {Gilbreth}}, \bibinfo {author} {\bibfnamefont {J.}~\bibnamefont
  {Giles}}, \bibinfo {author} {\bibfnamefont {E.}~\bibnamefont {Glynn}},
  \bibinfo {author} {\bibfnamefont {A.}~\bibnamefont {Hall}}, \bibinfo {author}
  {\bibfnamefont {A.~M.}\ \bibnamefont {Hankin}}, \bibinfo {author}
  {\bibfnamefont {A.}~\bibnamefont {Hansen}}, \bibinfo {author} {\bibfnamefont
  {D.}~\bibnamefont {Hayes}}, \bibinfo {author} {\bibfnamefont
  {B.}~\bibnamefont {Higashi}}, \bibinfo {author} {\bibfnamefont {I.~M.}\
  \bibnamefont {Hoffman}}, \bibinfo {author} {\bibfnamefont {B.}~\bibnamefont
  {Horning}}, \bibinfo {author} {\bibfnamefont {J.~J.}\ \bibnamefont {Hout}},
  \bibinfo {author} {\bibfnamefont {R.}~\bibnamefont {Jacobs}}, \bibinfo
  {author} {\bibfnamefont {J.}~\bibnamefont {Johansen}}, \bibinfo {author}
  {\bibfnamefont {L.}~\bibnamefont {Jones}}, \bibinfo {author} {\bibfnamefont
  {J.}~\bibnamefont {Karcz}}, \bibinfo {author} {\bibfnamefont
  {T.}~\bibnamefont {Klein}}, \bibinfo {author} {\bibfnamefont
  {P.}~\bibnamefont {Lauria}}, \bibinfo {author} {\bibfnamefont
  {P.}~\bibnamefont {Lee}}, \bibinfo {author} {\bibfnamefont {D.}~\bibnamefont
  {Liefer}}, \bibinfo {author} {\bibfnamefont {S.~T.}\ \bibnamefont {Lu}},
  \bibinfo {author} {\bibfnamefont {D.}~\bibnamefont {Lucchetti}}, \bibinfo
  {author} {\bibfnamefont {C.}~\bibnamefont {Lytle}}, \bibinfo {author}
  {\bibfnamefont {A.}~\bibnamefont {Malm}}, \bibinfo {author} {\bibfnamefont
  {M.}~\bibnamefont {Matheny}}, \bibinfo {author} {\bibfnamefont
  {B.}~\bibnamefont {Mathewson}}, \bibinfo {author} {\bibfnamefont
  {K.}~\bibnamefont {Mayer}}, \bibinfo {author} {\bibfnamefont {D.~B.}\
  \bibnamefont {Miller}}, \bibinfo {author} {\bibfnamefont {M.}~\bibnamefont
  {Mills}}, \bibinfo {author} {\bibfnamefont {B.}~\bibnamefont {Neyenhuis}},
  \bibinfo {author} {\bibfnamefont {L.}~\bibnamefont {Nugent}}, \bibinfo
  {author} {\bibfnamefont {S.}~\bibnamefont {Olson}}, \bibinfo {author}
  {\bibfnamefont {J.}~\bibnamefont {Parks}}, \bibinfo {author} {\bibfnamefont
  {G.~N.}\ \bibnamefont {Price}}, \bibinfo {author} {\bibfnamefont
  {Z.}~\bibnamefont {Price}}, \bibinfo {author} {\bibfnamefont
  {M.}~\bibnamefont {Pugh}}, \bibinfo {author} {\bibfnamefont {A.}~\bibnamefont
  {Ransford}}, \bibinfo {author} {\bibfnamefont {A.~P.}\ \bibnamefont {Reed}},
  \bibinfo {author} {\bibfnamefont {C.}~\bibnamefont {Roman}}, \bibinfo
  {author} {\bibfnamefont {M.}~\bibnamefont {Rowe}}, \bibinfo {author}
  {\bibfnamefont {C.}~\bibnamefont {Ryan-Anderson}}, \bibinfo {author}
  {\bibfnamefont {S.}~\bibnamefont {Sanders}}, \bibinfo {author} {\bibfnamefont
  {J.}~\bibnamefont {Sedlacek}}, \bibinfo {author} {\bibfnamefont
  {P.}~\bibnamefont {Shevchuk}}, \bibinfo {author} {\bibfnamefont
  {P.}~\bibnamefont {Siegfried}}, \bibinfo {author} {\bibfnamefont
  {T.}~\bibnamefont {Skripka}}, \bibinfo {author} {\bibfnamefont
  {B.}~\bibnamefont {Spaun}}, \bibinfo {author} {\bibfnamefont {R.~T.}\
  \bibnamefont {Sprenkle}}, \bibinfo {author} {\bibfnamefont {R.~P.}\
  \bibnamefont {Stutz}}, \bibinfo {author} {\bibfnamefont {M.}~\bibnamefont
  {Swallows}}, \bibinfo {author} {\bibfnamefont {R.~I.}\ \bibnamefont {Tobey}},
  \bibinfo {author} {\bibfnamefont {A.}~\bibnamefont {Tran}}, \bibinfo {author}
  {\bibfnamefont {T.}~\bibnamefont {Tran}}, \bibinfo {author} {\bibfnamefont
  {E.}~\bibnamefont {Vogt}}, \bibinfo {author} {\bibfnamefont {C.}~\bibnamefont
  {Volin}}, \bibinfo {author} {\bibfnamefont {J.}~\bibnamefont {Walker}},
  \bibinfo {author} {\bibfnamefont {A.~M.}\ \bibnamefont {Zolot}},\ and\
  \bibinfo {author} {\bibfnamefont {J.~M.}\ \bibnamefont {Pino}},\ }\bibfield
  {title} {\bibinfo {title} {A race-track trapped-ion quantum processor},\
  }\href {https://doi.org/10.1103/physrevx.13.041052} {\bibfield  {journal}
  {\bibinfo  {journal} {Physical Review X}\ }\textbf {\bibinfo {volume} {13}}
  (\bibinfo {year} {2023})}\BibitemShut {NoStop}%
\bibitem [{\citenamefont {Bluvstein}\ \emph {et~al.}(2023)\citenamefont
  {Bluvstein}, \citenamefont {Evered}, \citenamefont {Geim}, \citenamefont
  {Li}, \citenamefont {Zhou}, \citenamefont {Manovitz}, \citenamefont {Ebadi},
  \citenamefont {Cain}, \citenamefont {Kalinowski}, \citenamefont {Hangleiter},
  \citenamefont {Bonilla~Ataides}, \citenamefont {Maskara}, \citenamefont
  {Cong}, \citenamefont {Gao}, \citenamefont {Sales~Rodriguez}, \citenamefont
  {Karolyshyn}, \citenamefont {Semeghini}, \citenamefont {Gullans},
  \citenamefont {Greiner}, \citenamefont {Vuleti\'{c}},\ and\ \citenamefont
  {Lukin}}]{Bluvstein2023}%
  \BibitemOpen
  \bibfield  {author} {\bibinfo {author} {\bibfnamefont {D.}~\bibnamefont
  {Bluvstein}}, \bibinfo {author} {\bibfnamefont {S.~J.}\ \bibnamefont
  {Evered}}, \bibinfo {author} {\bibfnamefont {A.~A.}\ \bibnamefont {Geim}},
  \bibinfo {author} {\bibfnamefont {S.~H.}\ \bibnamefont {Li}}, \bibinfo
  {author} {\bibfnamefont {H.}~\bibnamefont {Zhou}}, \bibinfo {author}
  {\bibfnamefont {T.}~\bibnamefont {Manovitz}}, \bibinfo {author}
  {\bibfnamefont {S.}~\bibnamefont {Ebadi}}, \bibinfo {author} {\bibfnamefont
  {M.}~\bibnamefont {Cain}}, \bibinfo {author} {\bibfnamefont {M.}~\bibnamefont
  {Kalinowski}}, \bibinfo {author} {\bibfnamefont {D.}~\bibnamefont
  {Hangleiter}}, \bibinfo {author} {\bibfnamefont {J.~P.}\ \bibnamefont
  {Bonilla~Ataides}}, \bibinfo {author} {\bibfnamefont {N.}~\bibnamefont
  {Maskara}}, \bibinfo {author} {\bibfnamefont {I.}~\bibnamefont {Cong}},
  \bibinfo {author} {\bibfnamefont {X.}~\bibnamefont {Gao}}, \bibinfo {author}
  {\bibfnamefont {P.}~\bibnamefont {Sales~Rodriguez}}, \bibinfo {author}
  {\bibfnamefont {T.}~\bibnamefont {Karolyshyn}}, \bibinfo {author}
  {\bibfnamefont {G.}~\bibnamefont {Semeghini}}, \bibinfo {author}
  {\bibfnamefont {M.~J.}\ \bibnamefont {Gullans}}, \bibinfo {author}
  {\bibfnamefont {M.}~\bibnamefont {Greiner}}, \bibinfo {author} {\bibfnamefont
  {V.}~\bibnamefont {Vuleti\'{c}}},\ and\ \bibinfo {author} {\bibfnamefont
  {M.~D.}\ \bibnamefont {Lukin}},\ }\bibfield  {title} {\bibinfo {title}
  {Logical quantum processor based on reconfigurable atom arrays},\ }\href
  {https://doi.org/10.1038/s41586-023-06927-3} {\bibfield  {journal} {\bibinfo
  {journal} {Nature}\ }\textbf {\bibinfo {volume} {626}},\ \bibinfo {pages}
  {58} (\bibinfo {year} {2023})}\BibitemShut {NoStop}%
\bibitem [{\citenamefont {DeCross}\ \emph {et~al.}(2025)\citenamefont
  {DeCross}, \citenamefont {Haghshenas}, \citenamefont {Liu}, \citenamefont
  {Alexeev}, \citenamefont {Baldwin}, \citenamefont {Bartolotta}, \citenamefont
  {Bohn}, \citenamefont {Chertkov}, \citenamefont {Colina}, \citenamefont
  {DelVento}, \citenamefont {Dreiling}, \citenamefont {Foltz}, \citenamefont
  {Gaebler}, \citenamefont {Gatterman}, \citenamefont {Gilbreth}, \citenamefont
  {Gray}, \citenamefont {Gresh}, \citenamefont {Hewitt}, \citenamefont
  {Hutson}, \citenamefont {Johansen}, \citenamefont {Lucchetti}, \citenamefont
  {Lykov}, \citenamefont {Madjarov}, \citenamefont {Mayer}, \citenamefont
  {Mills}, \citenamefont {Niroula}, \citenamefont {Rinaldi}, \citenamefont
  {Siegfried}, \citenamefont {Tiemann}, \citenamefont {Volin}, \citenamefont
  {Walker}, \citenamefont {Shaydulin}, \citenamefont {Pistoia}, \citenamefont
  {Moses}, \citenamefont {Hayes}, \citenamefont {Neyenhuis}, \citenamefont
  {Stutz},\ and\ \citenamefont {Foss-Feig}}]{2406.02501}%
  \BibitemOpen
  \bibfield  {author} {\bibinfo {author} {\bibfnamefont {M.}~\bibnamefont
  {DeCross}}, \bibinfo {author} {\bibfnamefont {R.}~\bibnamefont {Haghshenas}},
  \bibinfo {author} {\bibfnamefont {M.}~\bibnamefont {Liu}}, \bibinfo {author}
  {\bibfnamefont {Y.}~\bibnamefont {Alexeev}}, \bibinfo {author} {\bibfnamefont
  {C.~H.}\ \bibnamefont {Baldwin}}, \bibinfo {author} {\bibfnamefont {J.~P.}\
  \bibnamefont {Bartolotta}}, \bibinfo {author} {\bibfnamefont
  {M.}~\bibnamefont {Bohn}}, \bibinfo {author} {\bibfnamefont {E.}~\bibnamefont
  {Chertkov}}, \bibinfo {author} {\bibfnamefont {J.}~\bibnamefont {Colina}},
  \bibinfo {author} {\bibfnamefont {D.}~\bibnamefont {DelVento}}, \bibinfo
  {author} {\bibfnamefont {J.~M.}\ \bibnamefont {Dreiling}}, \bibinfo {author}
  {\bibfnamefont {C.}~\bibnamefont {Foltz}}, \bibinfo {author} {\bibfnamefont
  {J.~P.}\ \bibnamefont {Gaebler}}, \bibinfo {author} {\bibfnamefont {T.~M.}\
  \bibnamefont {Gatterman}}, \bibinfo {author} {\bibfnamefont {C.~N.}\
  \bibnamefont {Gilbreth}}, \bibinfo {author} {\bibfnamefont {J.}~\bibnamefont
  {Gray}}, \bibinfo {author} {\bibfnamefont {D.}~\bibnamefont {Gresh}},
  \bibinfo {author} {\bibfnamefont {N.}~\bibnamefont {Hewitt}}, \bibinfo
  {author} {\bibfnamefont {R.~B.}\ \bibnamefont {Hutson}}, \bibinfo {author}
  {\bibfnamefont {J.}~\bibnamefont {Johansen}}, \bibinfo {author}
  {\bibfnamefont {D.}~\bibnamefont {Lucchetti}}, \bibinfo {author}
  {\bibfnamefont {D.}~\bibnamefont {Lykov}}, \bibinfo {author} {\bibfnamefont
  {I.~S.}\ \bibnamefont {Madjarov}}, \bibinfo {author} {\bibfnamefont
  {K.}~\bibnamefont {Mayer}}, \bibinfo {author} {\bibfnamefont
  {M.}~\bibnamefont {Mills}}, \bibinfo {author} {\bibfnamefont
  {P.}~\bibnamefont {Niroula}}, \bibinfo {author} {\bibfnamefont
  {E.}~\bibnamefont {Rinaldi}}, \bibinfo {author} {\bibfnamefont {P.~E.}\
  \bibnamefont {Siegfried}}, \bibinfo {author} {\bibfnamefont {B.~G.}\
  \bibnamefont {Tiemann}}, \bibinfo {author} {\bibfnamefont {C.}~\bibnamefont
  {Volin}}, \bibinfo {author} {\bibfnamefont {J.}~\bibnamefont {Walker}},
  \bibinfo {author} {\bibfnamefont {R.}~\bibnamefont {Shaydulin}}, \bibinfo
  {author} {\bibfnamefont {M.}~\bibnamefont {Pistoia}}, \bibinfo {author}
  {\bibfnamefont {S.~A.}\ \bibnamefont {Moses}}, \bibinfo {author}
  {\bibfnamefont {D.}~\bibnamefont {Hayes}}, \bibinfo {author} {\bibfnamefont
  {B.}~\bibnamefont {Neyenhuis}}, \bibinfo {author} {\bibfnamefont {R.~P.}\
  \bibnamefont {Stutz}},\ and\ \bibinfo {author} {\bibfnamefont
  {M.}~\bibnamefont {Foss-Feig}},\ }\bibfield  {title} {\bibinfo {title}
  {Computational power of random quantum circuits in arbitrary geometries},\
  }\href {https://doi.org/10.1103/PhysRevX.15.021052} {\bibfield  {journal}
  {\bibinfo  {journal} {Physical Review X}\ }\textbf {\bibinfo {volume} {15}},\
  \bibinfo {pages} {021052} (\bibinfo {year} {2025})}\BibitemShut {NoStop}%
\bibitem [{\citenamefont {Abbas}\ \emph {et~al.}(2023)\citenamefont {Abbas},
  \citenamefont {King}, \citenamefont {Huang}, \citenamefont {Huggins},
  \citenamefont {Movassagh}, \citenamefont {Gilboa},\ and\ \citenamefont
  {McClean}}]{2305.13362}%
  \BibitemOpen
  \bibfield  {author} {\bibinfo {author} {\bibfnamefont {A.}~\bibnamefont
  {Abbas}}, \bibinfo {author} {\bibfnamefont {R.}~\bibnamefont {King}},
  \bibinfo {author} {\bibfnamefont {H.-Y.}\ \bibnamefont {Huang}}, \bibinfo
  {author} {\bibfnamefont {W.~J.}\ \bibnamefont {Huggins}}, \bibinfo {author}
  {\bibfnamefont {R.}~\bibnamefont {Movassagh}}, \bibinfo {author}
  {\bibfnamefont {D.}~\bibnamefont {Gilboa}},\ and\ \bibinfo {author}
  {\bibfnamefont {J.~R.}\ \bibnamefont {McClean}},\ }\bibfield  {title}
  {\bibinfo {title} {On quantum backpropagation, information reuse, and
  cheating measurement collapse},\ }\href
  {https://doi.org/10.48550/arXiv.2305.13362} {\bibfield  {journal} {\bibinfo
  {journal} {arXiv:2305.13362}\ } (\bibinfo {year} {2023})}\BibitemShut
  {NoStop}%
\bibitem [{\citenamefont {Larson}\ \emph {et~al.}(2019)\citenamefont {Larson},
  \citenamefont {Menickelly},\ and\ \citenamefont {Wild}}]{LMW2019}%
  \BibitemOpen
  \bibfield  {author} {\bibinfo {author} {\bibfnamefont {J.}~\bibnamefont
  {Larson}}, \bibinfo {author} {\bibfnamefont {M.}~\bibnamefont {Menickelly}},\
  and\ \bibinfo {author} {\bibfnamefont {S.~M.}\ \bibnamefont {Wild}},\
  }\bibfield  {title} {\bibinfo {title} {Derivative-free optimization
  methods},\ }\href {https://doi.org/10.1017/s0962492919000060} {\bibfield
  {journal} {\bibinfo  {journal} {Acta Numerica}\ }\textbf {\bibinfo {volume}
  {28}},\ \bibinfo {pages} {287} (\bibinfo {year} {2019})}\BibitemShut
  {NoStop}%
\bibitem [{\citenamefont {Larson}\ and\ \citenamefont
  {Billups}(2016)}]{Larson2016}%
  \BibitemOpen
  \bibfield  {author} {\bibinfo {author} {\bibfnamefont {J.}~\bibnamefont
  {Larson}}\ and\ \bibinfo {author} {\bibfnamefont {S.~C.}\ \bibnamefont
  {Billups}},\ }\bibfield  {title} {\bibinfo {title} {Stochastic
  derivative-free optimization using a trust region framework},\ }\href
  {https://doi.org/10.1007/s10589-016-9827-z} {\bibfield  {journal} {\bibinfo
  {journal} {Computational Optimization and Applications}\ }\textbf {\bibinfo
  {volume} {64}},\ \bibinfo {pages} {619} (\bibinfo {year} {2016})}\BibitemShut
  {NoStop}%
\bibitem [{\citenamefont {Chen}\ \emph {et~al.}(2018)\citenamefont {Chen},
  \citenamefont {Menickelly},\ and\ \citenamefont
  {Scheinberg}}]{stormoriginal}%
  \BibitemOpen
  \bibfield  {author} {\bibinfo {author} {\bibfnamefont {R.}~\bibnamefont
  {Chen}}, \bibinfo {author} {\bibfnamefont {M.}~\bibnamefont {Menickelly}},\
  and\ \bibinfo {author} {\bibfnamefont {K.}~\bibnamefont {Scheinberg}},\
  }\bibfield  {title} {\bibinfo {title} {Stochastic optimization using a
  trust-region method and random models},\ }\href
  {https://doi.org/10.1007/s10107-017-1141-8} {\bibfield  {journal} {\bibinfo
  {journal} {Mathematical Programming}\ }\textbf {\bibinfo {volume} {169}},\
  \bibinfo {pages} {447} (\bibinfo {year} {2018})}\BibitemShut {NoStop}%
\bibitem [{\citenamefont {Shashaani}\ \emph {et~al.}(2018)\citenamefont
  {Shashaani}, \citenamefont {Hashemi},\ and\ \citenamefont
  {Pasupathy}}]{Shashaani2018}%
  \BibitemOpen
  \bibfield  {author} {\bibinfo {author} {\bibfnamefont {S.}~\bibnamefont
  {Shashaani}}, \bibinfo {author} {\bibfnamefont {F.~S.}\ \bibnamefont
  {Hashemi}},\ and\ \bibinfo {author} {\bibfnamefont {R.}~\bibnamefont
  {Pasupathy}},\ }\bibfield  {title} {\bibinfo {title} {{ASTRO}-{DF}: {A} class
  of adaptive sampling trust-region algorithms for derivative-free stochastic
  optimization},\ }\href {https://doi.org/10.1137/15m1042425} {\bibfield
  {journal} {\bibinfo  {journal} {{SIAM} Journal on Optimization}\ }\textbf
  {\bibinfo {volume} {28}},\ \bibinfo {pages} {3145} (\bibinfo {year}
  {2018})}\BibitemShut {NoStop}%
\bibitem [{\citenamefont {Scriva}\ \emph {et~al.}(2024)\citenamefont {Scriva},
  \citenamefont {Astrakhantsev}, \citenamefont {Pilati},\ and\ \citenamefont
  {Mazzola}}]{scriva2024challenges}%
  \BibitemOpen
  \bibfield  {author} {\bibinfo {author} {\bibfnamefont {G.}~\bibnamefont
  {Scriva}}, \bibinfo {author} {\bibfnamefont {N.}~\bibnamefont
  {Astrakhantsev}}, \bibinfo {author} {\bibfnamefont {S.}~\bibnamefont
  {Pilati}},\ and\ \bibinfo {author} {\bibfnamefont {G.}~\bibnamefont
  {Mazzola}},\ }\bibfield  {title} {\bibinfo {title} {Challenges of variational
  quantum optimization with measurement shot noise},\ }\href
  {https://doi.org/10.1103/PhysRevA.109.032408} {\bibfield  {journal} {\bibinfo
   {journal} {Phys. Rev. A}\ }\textbf {\bibinfo {volume} {109}},\ \bibinfo
  {pages} {032408} (\bibinfo {year} {2024})}\BibitemShut {NoStop}%
\bibitem [{\citenamefont {Sachdeva}\ \emph {et~al.}(2024)\citenamefont
  {Sachdeva}, \citenamefont {Harnett}, \citenamefont {Maity}, \citenamefont
  {Marsh}, \citenamefont {Wang}, \citenamefont {Winick}, \citenamefont
  {Dougherty}, \citenamefont {Canuto}, \citenamefont {Chong}, \citenamefont
  {Hush} \emph {et~al.}}]{sachdeva2024quantum}%
  \BibitemOpen
  \bibfield  {author} {\bibinfo {author} {\bibfnamefont {N.}~\bibnamefont
  {Sachdeva}}, \bibinfo {author} {\bibfnamefont {G.~S.}\ \bibnamefont
  {Harnett}}, \bibinfo {author} {\bibfnamefont {S.}~\bibnamefont {Maity}},
  \bibinfo {author} {\bibfnamefont {S.}~\bibnamefont {Marsh}}, \bibinfo
  {author} {\bibfnamefont {Y.}~\bibnamefont {Wang}}, \bibinfo {author}
  {\bibfnamefont {A.}~\bibnamefont {Winick}}, \bibinfo {author} {\bibfnamefont
  {R.}~\bibnamefont {Dougherty}}, \bibinfo {author} {\bibfnamefont
  {D.}~\bibnamefont {Canuto}}, \bibinfo {author} {\bibfnamefont {Y.~Q.}\
  \bibnamefont {Chong}}, \bibinfo {author} {\bibfnamefont {M.}~\bibnamefont
  {Hush}}, \emph {et~al.},\ }\bibfield  {title} {\bibinfo {title} {Quantum
  optimization using a 127-qubit gate-model {IBM} quantum computer can
  outperform quantum annealers for nontrivial binary optimization problems},\
  }\href {https://doi.org/10.48550/arXiv.2406.01743} {\bibfield  {journal}
  {\bibinfo  {journal} {arXiv:2406.01743}\ } (\bibinfo {year}
  {2024})}\BibitemShut {NoStop}%
\bibitem [{\citenamefont {Menickelly}\ \emph {et~al.}(2023)\citenamefont
  {Menickelly}, \citenamefont {Ha},\ and\ \citenamefont
  {Otten}}]{menickelly2023latency}%
  \BibitemOpen
  \bibfield  {author} {\bibinfo {author} {\bibfnamefont {M.}~\bibnamefont
  {Menickelly}}, \bibinfo {author} {\bibfnamefont {Y.}~\bibnamefont {Ha}},\
  and\ \bibinfo {author} {\bibfnamefont {M.}~\bibnamefont {Otten}},\ }\bibfield
   {title} {\bibinfo {title} {Latency considerations for stochastic optimizers
  in variational quantum algorithms},\ }\href
  {https://doi.org/10.22331/q-2023-03-16-949} {\bibfield  {journal} {\bibinfo
  {journal} {Quantum}\ }\textbf {\bibinfo {volume} {7}},\ \bibinfo {pages}
  {949} (\bibinfo {year} {2023})}\BibitemShut {NoStop}%
\bibitem [{\citenamefont {Ito}(2023)}]{ito2023latency}%
  \BibitemOpen
  \bibfield  {author} {\bibinfo {author} {\bibfnamefont {K.}~\bibnamefont
  {Ito}},\ }\bibfield  {title} {\bibinfo {title} {Latency-aware adaptive shot
  allocation for run-time efficient variational quantum algorithms},\ }\href
  {https://doi.org/10.48550/arXiv.2302.04422} {\bibfield  {journal} {\bibinfo
  {journal} {arXiv:2302.04422}\ } (\bibinfo {year} {2023})}\BibitemShut
  {NoStop}%
\bibitem [{\citenamefont {Gu}\ \emph {et~al.}(2021)\citenamefont {Gu},
  \citenamefont {Lowe}, \citenamefont {Dub}, \citenamefont {Coles},\ and\
  \citenamefont {Arrasmith}}]{gu2021adaptive}%
  \BibitemOpen
  \bibfield  {author} {\bibinfo {author} {\bibfnamefont {A.}~\bibnamefont
  {Gu}}, \bibinfo {author} {\bibfnamefont {A.}~\bibnamefont {Lowe}}, \bibinfo
  {author} {\bibfnamefont {P.~A.}\ \bibnamefont {Dub}}, \bibinfo {author}
  {\bibfnamefont {P.~J.}\ \bibnamefont {Coles}},\ and\ \bibinfo {author}
  {\bibfnamefont {A.}~\bibnamefont {Arrasmith}},\ }\bibfield  {title} {\bibinfo
  {title} {Adaptive shot allocation for fast convergence in variational quantum
  algorithms},\ }\href {https://doi.org/10.48550/arXiv.2108.10434} {\bibfield
  {journal} {\bibinfo  {journal} {arXiv:2108.10434}\ } (\bibinfo {year}
  {2021})}\BibitemShut {NoStop}%
\bibitem [{\citenamefont {Arrasmith}\ \emph {et~al.}(2020)\citenamefont
  {Arrasmith}, \citenamefont {Cincio}, \citenamefont {Somma},\ and\
  \citenamefont {Coles}}]{arrasmith2020operator}%
  \BibitemOpen
  \bibfield  {author} {\bibinfo {author} {\bibfnamefont {A.}~\bibnamefont
  {Arrasmith}}, \bibinfo {author} {\bibfnamefont {L.}~\bibnamefont {Cincio}},
  \bibinfo {author} {\bibfnamefont {R.~D.}\ \bibnamefont {Somma}},\ and\
  \bibinfo {author} {\bibfnamefont {P.~J.}\ \bibnamefont {Coles}},\ }\bibfield
  {title} {\bibinfo {title} {Operator sampling for shot-frugal optimization in
  variational algorithms},\ }\href {https://doi.org/10.48550/arXiv.2004.06252}
  {\bibfield  {journal} {\bibinfo  {journal} {arXiv:2004.06252}\ } (\bibinfo
  {year} {2020})}\BibitemShut {NoStop}%
\bibitem [{\citenamefont {Sung}\ \emph {et~al.}(2020)\citenamefont {Sung},
  \citenamefont {Yao}, \citenamefont {Harrigan}, \citenamefont {Rubin},
  \citenamefont {Jiang}, \citenamefont {Lin}, \citenamefont {Babbush},\ and\
  \citenamefont {McClean}}]{Sung2020}%
  \BibitemOpen
  \bibfield  {author} {\bibinfo {author} {\bibfnamefont {K.~J.}\ \bibnamefont
  {Sung}}, \bibinfo {author} {\bibfnamefont {J.}~\bibnamefont {Yao}}, \bibinfo
  {author} {\bibfnamefont {M.~P.}\ \bibnamefont {Harrigan}}, \bibinfo {author}
  {\bibfnamefont {N.~C.}\ \bibnamefont {Rubin}}, \bibinfo {author}
  {\bibfnamefont {Z.}~\bibnamefont {Jiang}}, \bibinfo {author} {\bibfnamefont
  {L.}~\bibnamefont {Lin}}, \bibinfo {author} {\bibfnamefont {R.}~\bibnamefont
  {Babbush}},\ and\ \bibinfo {author} {\bibfnamefont {J.~R.}\ \bibnamefont
  {McClean}},\ }\bibfield  {title} {\bibinfo {title} {Using models to improve
  optimizers for variational quantum algorithms},\ }\href
  {https://doi.org/10.1088/2058-9565/abb6d9} {\bibfield  {journal} {\bibinfo
  {journal} {Quantum Science and Technology}\ }\textbf {\bibinfo {volume}
  {5}},\ \bibinfo {pages} {044008} (\bibinfo {year} {2020})}\BibitemShut
  {NoStop}%
\bibitem [{\citenamefont {Zhu}\ \emph {et~al.}(2024)\citenamefont {Zhu},
  \citenamefont {Liang}, \citenamefont {Yang},\ and\ \citenamefont
  {Li}}]{zhu2023optimizing}%
  \BibitemOpen
  \bibfield  {author} {\bibinfo {author} {\bibfnamefont {L.}~\bibnamefont
  {Zhu}}, \bibinfo {author} {\bibfnamefont {S.}~\bibnamefont {Liang}}, \bibinfo
  {author} {\bibfnamefont {C.}~\bibnamefont {Yang}},\ and\ \bibinfo {author}
  {\bibfnamefont {X.}~\bibnamefont {Li}},\ }\bibfield  {title} {\bibinfo
  {title} {Optimizing shot assignment in variational quantum eigensolver
  measurement},\ }\href {https://doi.org/10.1021/acs.jctc.3c01113} {\bibfield
  {journal} {\bibinfo  {journal} {Journal of Chemical Theory and Computation}\
  }\textbf {\bibinfo {volume} {20}},\ \bibinfo {pages} {2390} (\bibinfo {year}
  {2024})}\BibitemShut {NoStop}%
\bibitem [{\citenamefont {Moussa}\ \emph {et~al.}(2023)\citenamefont {Moussa},
  \citenamefont {Gordon}, \citenamefont {Baczyk}, \citenamefont {Cerezo},
  \citenamefont {Cincio},\ and\ \citenamefont {Coles}}]{moussa2023resource}%
  \BibitemOpen
  \bibfield  {author} {\bibinfo {author} {\bibfnamefont {C.}~\bibnamefont
  {Moussa}}, \bibinfo {author} {\bibfnamefont {M.~H.}\ \bibnamefont {Gordon}},
  \bibinfo {author} {\bibfnamefont {M.}~\bibnamefont {Baczyk}}, \bibinfo
  {author} {\bibfnamefont {M.}~\bibnamefont {Cerezo}}, \bibinfo {author}
  {\bibfnamefont {L.}~\bibnamefont {Cincio}},\ and\ \bibinfo {author}
  {\bibfnamefont {P.~J.}\ \bibnamefont {Coles}},\ }\bibfield  {title} {\bibinfo
  {title} {Resource frugal optimizer for quantum machine learning},\ }\href
  {https://doi.org/10.1088/2058-9565/acef55} {\bibfield  {journal} {\bibinfo
  {journal} {Quantum Science and Technology}\ }\textbf {\bibinfo {volume}
  {8}},\ \bibinfo {pages} {045019} (\bibinfo {year} {2023})}\BibitemShut
  {NoStop}%
\bibitem [{\citenamefont {K{\"u}bler}\ \emph {et~al.}(2020)\citenamefont
  {K{\"u}bler}, \citenamefont {Arrasmith}, \citenamefont {Cincio},\ and\
  \citenamefont {Coles}}]{kubler2020adaptive}%
  \BibitemOpen
  \bibfield  {author} {\bibinfo {author} {\bibfnamefont {J.~M.}\ \bibnamefont
  {K{\"u}bler}}, \bibinfo {author} {\bibfnamefont {A.}~\bibnamefont
  {Arrasmith}}, \bibinfo {author} {\bibfnamefont {L.}~\bibnamefont {Cincio}},\
  and\ \bibinfo {author} {\bibfnamefont {P.~J.}\ \bibnamefont {Coles}},\
  }\bibfield  {title} {\bibinfo {title} {An adaptive optimizer for
  measurement-frugal variational algorithms},\ }\href
  {https://doi.org/10.22331/q-2020-05-11-263} {\bibfield  {journal} {\bibinfo
  {journal} {Quantum}\ }\textbf {\bibinfo {volume} {4}},\ \bibinfo {pages}
  {263} (\bibinfo {year} {2020})}\BibitemShut {NoStop}%
\bibitem [{\citenamefont {Polloreno}\ and\ \citenamefont
  {Smith}(2022)}]{polloreno2022qaoa}%
  \BibitemOpen
  \bibfield  {author} {\bibinfo {author} {\bibfnamefont {A.~M.}\ \bibnamefont
  {Polloreno}}\ and\ \bibinfo {author} {\bibfnamefont {G.}~\bibnamefont
  {Smith}},\ }\bibfield  {title} {\bibinfo {title} {The {QAOA} with slow
  measurements},\ }\href {https://doi.org/10.48550/arXiv.2205.06845} {\bibfield
   {journal} {\bibinfo  {journal} {arXiv:2205.06845}\ } (\bibinfo {year}
  {2022})}\BibitemShut {NoStop}%
\bibitem [{\citenamefont {Cheng}\ \emph {et~al.}(2024)\citenamefont {Cheng},
  \citenamefont {Chen}, \citenamefont {Zhang},\ and\ \citenamefont
  {Zhang}}]{cheng2024quantum}%
  \BibitemOpen
  \bibfield  {author} {\bibinfo {author} {\bibfnamefont {L.}~\bibnamefont
  {Cheng}}, \bibinfo {author} {\bibfnamefont {Y.-Q.}\ \bibnamefont {Chen}},
  \bibinfo {author} {\bibfnamefont {S.-X.}\ \bibnamefont {Zhang}},\ and\
  \bibinfo {author} {\bibfnamefont {S.}~\bibnamefont {Zhang}},\ }\bibfield
  {title} {\bibinfo {title} {Quantum approximate optimization via
  learning-based adaptive optimization},\ }\href
  {https://doi.org/10.1038/s42005-024-01577-x} {\bibfield  {journal} {\bibinfo
  {journal} {Communications Physics}\ }\textbf {\bibinfo {volume} {7}},\
  \bibinfo {pages} {83} (\bibinfo {year} {2024})}\BibitemShut {NoStop}%
\bibitem [{\citenamefont {Billingsley}(2017)}]{billingsley2017probability}%
  \BibitemOpen
  \bibfield  {author} {\bibinfo {author} {\bibfnamefont {P.}~\bibnamefont
  {Billingsley}},\ }\href
  {https://www.wiley.com/en-us/Probability+and+Measure%2C+Anniversary+Edition-p-9781118122372}
  {\emph {\bibinfo {title} {Probability and Measure}}}\ (\bibinfo  {publisher}
  {John Wiley \& Sons},\ \bibinfo {year} {2017})\BibitemShut {NoStop}%
\bibitem [{\citenamefont {Larson}\ \emph {et~al.}(2024)\citenamefont {Larson},
  \citenamefont {Menickelly},\ and\ \citenamefont {Shi}}]{Shi2024}%
  \BibitemOpen
  \bibfield  {author} {\bibinfo {author} {\bibfnamefont {J.}~\bibnamefont
  {Larson}}, \bibinfo {author} {\bibfnamefont {M.}~\bibnamefont {Menickelly}},\
  and\ \bibinfo {author} {\bibfnamefont {J.}~\bibnamefont {Shi}},\ }\bibfield
  {title} {\bibinfo {title} {A novel noise-aware classical optimizer for
  variational quantum algorithms},\ }\href
  {https://doi.org/10.48550/arXiv.2401.10121} {\bibfield  {journal} {\bibinfo
  {journal} {arXiv:2401.10121}\ } (\bibinfo {year} {2024})}\BibitemShut
  {NoStop}%
\bibitem [{\citenamefont {B{\"a}rtschi}\ and\ \citenamefont
  {Eidenbenz}(2019)}]{bartschi2019deterministic}%
  \BibitemOpen
  \bibfield  {author} {\bibinfo {author} {\bibfnamefont {A.}~\bibnamefont
  {B{\"a}rtschi}}\ and\ \bibinfo {author} {\bibfnamefont {S.}~\bibnamefont
  {Eidenbenz}},\ }\bibfield  {title} {\bibinfo {title} {Deterministic
  preparation of {Dicke} states},\ }in\ \href
  {https://doi.org/10.1007/978-3-030-25027-0_9} {\emph {\bibinfo {booktitle}
  {International Symposium on Fundamentals of Computation Theory}}}\ (\bibinfo
  {organization} {Springer},\ \bibinfo {year} {2019})\ pp.\ \bibinfo {pages}
  {126--139}\BibitemShut {NoStop}%
\bibitem [{\citenamefont {Powell}(1994)}]{COBYLA}%
  \BibitemOpen
  \bibfield  {author} {\bibinfo {author} {\bibfnamefont {M.~J.~D.}\
  \bibnamefont {Powell}},\ }\bibfield  {title} {\bibinfo {title} {A direct
  search optimization method that models the objective and constraint functions
  by linear interpolation},\ }in\ \href
  {https://doi.org/10.1007/978-94-015-8330-5_4} {\emph {\bibinfo {booktitle}
  {Advances in Optimization and Numerical Analysis}}},\ \bibinfo {series}
  {Mathematics and Its Applications}, Vol.\ \bibinfo {volume} {275},\ \bibinfo
  {editor} {edited by\ \bibinfo {editor} {\bibfnamefont {S.}~\bibnamefont
  {Gomez}}\ and\ \bibinfo {editor} {\bibfnamefont {J.-P.}\ \bibnamefont
  {Hennart}}}\ (\bibinfo  {publisher} {Springer},\ \bibinfo {year} {1994})\
  pp.\ \bibinfo {pages} {51--67}\BibitemShut {NoStop}%
\bibitem [{\citenamefont {Powell}(2009)}]{BOBYQA}%
  \BibitemOpen
  \bibfield  {author} {\bibinfo {author} {\bibfnamefont {M.~J.~D.}\
  \bibnamefont {Powell}},\ }\href
  {http://www.damtp.cam.ac.uk/user/na/NA_papers/NA2009_06.pdf} {\emph {\bibinfo
  {title} {The {BOBYQA} algorithm for bound constrained optimization without
  derivatives}}},\ \bibinfo {type} {Tech. Rep.}\ \bibinfo {number} {NA2009/06}\
  (\bibinfo  {institution} {Department of Applied Mathematics and Theoretical
  Physics, Cambridge University},\ \bibinfo {address} {Cambridge, UK},\
  \bibinfo {year} {2009})\BibitemShut {NoStop}%
\bibitem [{\citenamefont {Nelder}\ and\ \citenamefont
  {Mead}(1965)}]{NELDERMEAD}%
  \BibitemOpen
  \bibfield  {author} {\bibinfo {author} {\bibfnamefont {J.~A.}\ \bibnamefont
  {Nelder}}\ and\ \bibinfo {author} {\bibfnamefont {R.}~\bibnamefont {Mead}},\
  }\bibfield  {title} {\bibinfo {title} {A simplex method for function
  minimization},\ }\href {https://doi.org/10.1093/comjnl/7.4.308} {\bibfield
  {journal} {\bibinfo  {journal} {The Computer Journal}\ }\textbf {\bibinfo
  {volume} {7}},\ \bibinfo {pages} {308} (\bibinfo {year} {1965})}\BibitemShut
  {NoStop}%
\bibitem [{\citenamefont {da~Silva~Santos}\ \emph {et~al.}(2010)\citenamefont
  {da~Silva~Santos}, \citenamefont {Goncalves},\ and\ \citenamefont
  {Hernandez-Figueroa}}]{ESCH}%
  \BibitemOpen
  \bibfield  {author} {\bibinfo {author} {\bibfnamefont {C.~H.}\ \bibnamefont
  {da~Silva~Santos}}, \bibinfo {author} {\bibfnamefont {M.~S.}\ \bibnamefont
  {Goncalves}},\ and\ \bibinfo {author} {\bibfnamefont {H.~E.}\ \bibnamefont
  {Hernandez-Figueroa}},\ }\bibfield  {title} {\bibinfo {title} {Designing
  novel photonic devices by bio-inspired computing},\ }\href
  {https://doi.org/10.1109/LPT.2010.2051222} {\bibfield  {journal} {\bibinfo
  {journal} {IEEE Photonics Technology Letters}\ }\textbf {\bibinfo {volume}
  {22}},\ \bibinfo {pages} {1177} (\bibinfo {year} {2010})}\BibitemShut
  {NoStop}%
\bibitem [{\citenamefont {Gablonsky}\ and\ \citenamefont
  {Kelley}(2001)}]{DIRECT_L}%
  \BibitemOpen
  \bibfield  {author} {\bibinfo {author} {\bibfnamefont {J.~M.}\ \bibnamefont
  {Gablonsky}}\ and\ \bibinfo {author} {\bibfnamefont {C.~T.}\ \bibnamefont
  {Kelley}},\ }\bibfield  {title} {\bibinfo {title} {A locally-biased form of
  the {DIRECT} algorithm},\ }\href {https://doi.org/10.1023/A:1017930332101}
  {\bibfield  {journal} {\bibinfo  {journal} {Journal of Global Optimization}\
  }\textbf {\bibinfo {volume} {21}},\ \bibinfo {pages} {27} (\bibinfo {year}
  {2001})}\BibitemShut {NoStop}%
\bibitem [{\citenamefont {Kaelo}\ and\ \citenamefont {Ali}(2006)}]{CRS2_LM}%
  \BibitemOpen
  \bibfield  {author} {\bibinfo {author} {\bibfnamefont {P.}~\bibnamefont
  {Kaelo}}\ and\ \bibinfo {author} {\bibfnamefont {M.}~\bibnamefont {Ali}},\
  }\bibfield  {title} {\bibinfo {title} {Some variants of the controlled random
  search algorithm for global optimization},\ }\href
  {https://doi.org/10.1007/s10957-006-9101-0} {\bibfield  {journal} {\bibinfo
  {journal} {Journal of Optimization Theory and Applications}\ }\textbf
  {\bibinfo {volume} {130}},\ \bibinfo {pages} {253} (\bibinfo {year}
  {2006})}\BibitemShut {NoStop}%
\bibitem [{\citenamefont {Spall}(1992)}]{SPSA}%
  \BibitemOpen
  \bibfield  {author} {\bibinfo {author} {\bibfnamefont {J.~C.}\ \bibnamefont
  {Spall}},\ }\bibfield  {title} {\bibinfo {title} {Multivariate stochastic
  approximation using a simultaneous perturbation gradient approximation},\
  }\href {https://doi.org/10.1109/9.119632} {\bibfield  {journal} {\bibinfo
  {journal} {IEEE Transactions on Automatic Control}\ }\textbf {\bibinfo
  {volume} {37}},\ \bibinfo {pages} {332} (\bibinfo {year} {1992})}\BibitemShut
  {NoStop}%
\bibitem [{\citenamefont {Berahas}\ \emph {et~al.}(2022)\citenamefont
  {Berahas}, \citenamefont {Cao}, \citenamefont {Choromanski},\ and\
  \citenamefont {Scheinberg}}]{GSLS}%
  \BibitemOpen
  \bibfield  {author} {\bibinfo {author} {\bibfnamefont {A.~S.}\ \bibnamefont
  {Berahas}}, \bibinfo {author} {\bibfnamefont {L.}~\bibnamefont {Cao}},
  \bibinfo {author} {\bibfnamefont {K.}~\bibnamefont {Choromanski}},\ and\
  \bibinfo {author} {\bibfnamefont {K.}~\bibnamefont {Scheinberg}},\ }\bibfield
   {title} {\bibinfo {title} {A theoretical and empirical comparison of
  gradient approximations in derivative-free optimization},\ }\href
  {https://doi.org/10.1007/s10208-021-09513-z} {\bibfield  {journal} {\bibinfo
  {journal} {Foundations of Computational Mathematics}\ }\textbf {\bibinfo
  {volume} {22}},\ \bibinfo {pages} {507} (\bibinfo {year} {2022})}\BibitemShut
  {NoStop}%
\bibitem [{\citenamefont {Kelley}(2011)}]{IMFIL}%
  \BibitemOpen
  \bibfield  {author} {\bibinfo {author} {\bibfnamefont {C.~T.}\ \bibnamefont
  {Kelley}},\ }\href {https://doi.org/10.1137/1.9781611971903} {\emph {\bibinfo
  {title} {Implicit Filtering}}}\ (\bibinfo  {publisher} {SIAM},\ \bibinfo
  {year} {2011})\BibitemShut {NoStop}%
\bibitem [{\citenamefont {Johnson}(2007)}]{NLopt}%
  \BibitemOpen
  \bibfield  {author} {\bibinfo {author} {\bibfnamefont {S.~G.}\ \bibnamefont
  {Johnson}},\ }\href@noop {} {\bibinfo {title} {The {NLopt}
  nonlinear-optimization package}},\ \bibinfo {howpublished}
  {\url{https://github.com/stevengj/nlopt}} (\bibinfo {year}
  {2007})\BibitemShut {NoStop}%
\bibitem [{\citenamefont {Cartis}\ \emph {et~al.}(2019)\citenamefont {Cartis},
  \citenamefont {Fiala}, \citenamefont {Marteau},\ and\ \citenamefont
  {Roberts}}]{cartis2019improving}%
  \BibitemOpen
  \bibfield  {author} {\bibinfo {author} {\bibfnamefont {C.}~\bibnamefont
  {Cartis}}, \bibinfo {author} {\bibfnamefont {J.}~\bibnamefont {Fiala}},
  \bibinfo {author} {\bibfnamefont {B.}~\bibnamefont {Marteau}},\ and\ \bibinfo
  {author} {\bibfnamefont {L.}~\bibnamefont {Roberts}},\ }\bibfield  {title}
  {\bibinfo {title} {Improving the flexibility and robustness of model-based
  derivative-free optimization solvers},\ }\href@noop {} {\bibfield  {journal}
  {\bibinfo  {journal} {ACM Transactions on Mathematical Software (TOMS)}\
  }\textbf {\bibinfo {volume} {45}},\ \bibinfo {pages} {1} (\bibinfo {year}
  {2019})}\BibitemShut {NoStop}%
\bibitem [{\citenamefont {Virtanen}\ \emph {et~al.}(2020)\citenamefont
  {Virtanen}, \citenamefont {Gommers}, \citenamefont {Oliphant}, \citenamefont
  {Haberland}, \citenamefont {Reddy}, \citenamefont {Cournapeau}, \citenamefont
  {Burovski}, \citenamefont {Peterson}, \citenamefont {Weckesser},
  \citenamefont {Bright}, \citenamefont {{van der Walt}}, \citenamefont
  {Brett}, \citenamefont {Wilson}, \citenamefont {Millman}, \citenamefont
  {Mayorov}, \citenamefont {Nelson}, \citenamefont {Jones}, \citenamefont
  {Kern}, \citenamefont {Larson}, \citenamefont {Carey}, \citenamefont {Polat},
  \citenamefont {Feng}, \citenamefont {Moore}, \citenamefont {{VanderPlas}},
  \citenamefont {Laxalde}, \citenamefont {Perktold}, \citenamefont {Cimrman},
  \citenamefont {Henriksen}, \citenamefont {Quintero}, \citenamefont {Harris},
  \citenamefont {Archibald}, \citenamefont {Ribeiro}, \citenamefont
  {Pedregosa}, \citenamefont {{van Mulbregt}},\ and\ \citenamefont {{SciPy 1.0
  Contributors}}}]{2020SciPy-NMeth}%
  \BibitemOpen
  \bibfield  {author} {\bibinfo {author} {\bibfnamefont {P.}~\bibnamefont
  {Virtanen}}, \bibinfo {author} {\bibfnamefont {R.}~\bibnamefont {Gommers}},
  \bibinfo {author} {\bibfnamefont {T.~E.}\ \bibnamefont {Oliphant}}, \bibinfo
  {author} {\bibfnamefont {M.}~\bibnamefont {Haberland}}, \bibinfo {author}
  {\bibfnamefont {T.}~\bibnamefont {Reddy}}, \bibinfo {author} {\bibfnamefont
  {D.}~\bibnamefont {Cournapeau}}, \bibinfo {author} {\bibfnamefont
  {E.}~\bibnamefont {Burovski}}, \bibinfo {author} {\bibfnamefont
  {P.}~\bibnamefont {Peterson}}, \bibinfo {author} {\bibfnamefont
  {W.}~\bibnamefont {Weckesser}}, \bibinfo {author} {\bibfnamefont
  {J.}~\bibnamefont {Bright}}, \bibinfo {author} {\bibfnamefont {S.~J.}\
  \bibnamefont {{van der Walt}}}, \bibinfo {author} {\bibfnamefont
  {M.}~\bibnamefont {Brett}}, \bibinfo {author} {\bibfnamefont
  {J.}~\bibnamefont {Wilson}}, \bibinfo {author} {\bibfnamefont {K.~J.}\
  \bibnamefont {Millman}}, \bibinfo {author} {\bibfnamefont {N.}~\bibnamefont
  {Mayorov}}, \bibinfo {author} {\bibfnamefont {A.~R.~J.}\ \bibnamefont
  {Nelson}}, \bibinfo {author} {\bibfnamefont {E.}~\bibnamefont {Jones}},
  \bibinfo {author} {\bibfnamefont {R.}~\bibnamefont {Kern}}, \bibinfo {author}
  {\bibfnamefont {E.}~\bibnamefont {Larson}}, \bibinfo {author} {\bibfnamefont
  {C.~J.}\ \bibnamefont {Carey}}, \bibinfo {author} {\bibfnamefont
  {{\.I}.}~\bibnamefont {Polat}}, \bibinfo {author} {\bibfnamefont
  {Y.}~\bibnamefont {Feng}}, \bibinfo {author} {\bibfnamefont {E.~W.}\
  \bibnamefont {Moore}}, \bibinfo {author} {\bibfnamefont {J.}~\bibnamefont
  {{VanderPlas}}}, \bibinfo {author} {\bibfnamefont {D.}~\bibnamefont
  {Laxalde}}, \bibinfo {author} {\bibfnamefont {J.}~\bibnamefont {Perktold}},
  \bibinfo {author} {\bibfnamefont {R.}~\bibnamefont {Cimrman}}, \bibinfo
  {author} {\bibfnamefont {I.}~\bibnamefont {Henriksen}}, \bibinfo {author}
  {\bibfnamefont {E.~A.}\ \bibnamefont {Quintero}}, \bibinfo {author}
  {\bibfnamefont {C.~R.}\ \bibnamefont {Harris}}, \bibinfo {author}
  {\bibfnamefont {A.~M.}\ \bibnamefont {Archibald}}, \bibinfo {author}
  {\bibfnamefont {A.~H.}\ \bibnamefont {Ribeiro}}, \bibinfo {author}
  {\bibfnamefont {F.}~\bibnamefont {Pedregosa}}, \bibinfo {author}
  {\bibfnamefont {P.}~\bibnamefont {{van Mulbregt}}},\ and\ \bibinfo {author}
  {\bibnamefont {{SciPy 1.0 Contributors}}},\ }\bibfield  {title} {\bibinfo
  {title} {{{SciPy} 1.0: Fundamental Algorithms for Scientific Computing in
  Python}},\ }\href {https://doi.org/10.1038/s41592-019-0686-2} {\bibfield
  {journal} {\bibinfo  {journal} {Nature Methods}\ }\textbf {\bibinfo {volume}
  {17}},\ \bibinfo {pages} {261} (\bibinfo {year} {2020})}\BibitemShut
  {NoStop}%
\bibitem [{\citenamefont {Ragonneau}\ and\ \citenamefont {Zhang}(2023)}]{PDFO}%
  \BibitemOpen
  \bibfield  {author} {\bibinfo {author} {\bibfnamefont {T.~M.}\ \bibnamefont
  {Ragonneau}}\ and\ \bibinfo {author} {\bibfnamefont {Z.}~\bibnamefont
  {Zhang}},\ }\bibfield  {title} {\bibinfo {title} {{PDFO}: A cross-platform
  package for {Powell}'s derivative-free optimization solvers},\ }\href
  {https://doi.org/10.48550/arXiv.2302.13246} {\bibfield  {journal} {\bibinfo
  {journal} {arXiv:2302.13246}\ } (\bibinfo {year} {2023})}\BibitemShut
  {NoStop}%
\bibitem [{\citenamefont {Lavrijsen}\ \emph {et~al.}(2020)\citenamefont
  {Lavrijsen}, \citenamefont {Tudor}, \citenamefont {M{\"u}ller}, \citenamefont
  {Iancu},\ and\ \citenamefont {De~Jong}}]{lavrijsen2020classical}%
  \BibitemOpen
  \bibfield  {author} {\bibinfo {author} {\bibfnamefont {W.}~\bibnamefont
  {Lavrijsen}}, \bibinfo {author} {\bibfnamefont {A.}~\bibnamefont {Tudor}},
  \bibinfo {author} {\bibfnamefont {J.}~\bibnamefont {M{\"u}ller}}, \bibinfo
  {author} {\bibfnamefont {C.}~\bibnamefont {Iancu}},\ and\ \bibinfo {author}
  {\bibfnamefont {W.}~\bibnamefont {De~Jong}},\ }\bibfield  {title} {\bibinfo
  {title} {Classical optimizers for noisy intermediate-scale quantum devices},\
  }in\ \href {https://doi.org/10.1109/QCE49297.2020.00041} {\emph {\bibinfo
  {booktitle} {International Conference on Quantum Computing and
  Engineering}}}\ (\bibinfo {organization} {IEEE},\ \bibinfo {year} {2020})\
  pp.\ \bibinfo {pages} {267--277}\BibitemShut {NoStop}%
\bibitem [{\citenamefont {Powell}(2002)}]{UOBYQA}%
  \BibitemOpen
  \bibfield  {author} {\bibinfo {author} {\bibfnamefont {M.~J.}\ \bibnamefont
  {Powell}},\ }\bibfield  {title} {\bibinfo {title} {{UOBYQA}: Unconstrained
  optimization by quadratic approximation},\ }\href
  {https://doi.org/10.1007/s101070100290} {\bibfield  {journal} {\bibinfo
  {journal} {Mathematical Programming}\ }\textbf {\bibinfo {volume} {92}},\
  \bibinfo {pages} {555} (\bibinfo {year} {2002})}\BibitemShut {NoStop}%
\bibitem [{\citenamefont {Powell}(2006)}]{NEWUOA}%
  \BibitemOpen
  \bibfield  {author} {\bibinfo {author} {\bibfnamefont {M.~J.~D.}\
  \bibnamefont {Powell}},\ }\bibfield  {title} {\bibinfo {title} {The {NEWUOA}
  software for unconstrained optimization without derivatives},\ }in\ \href
  {https://doi.org/10.1007/0-387-30065-1_16} {\emph {\bibinfo {booktitle}
  {Large-Scale Nonlinear Optimization}}},\ \bibinfo {series} {Nonconvex
  Optimization and Its Applications}, Vol.~\bibinfo {volume} {83},\ \bibinfo
  {editor} {edited by\ \bibinfo {editor} {\bibfnamefont {G.~D.}\ \bibnamefont
  {Pillo}}\ and\ \bibinfo {editor} {\bibfnamefont {M.}~\bibnamefont {Roma}}}\
  (\bibinfo  {publisher} {Springer},\ \bibinfo {year} {2006})\ pp.\ \bibinfo
  {pages} {255--297}\BibitemShut {NoStop}%
\bibitem [{\citenamefont {Powell}(2015)}]{Powell2015}%
  \BibitemOpen
  \bibfield  {author} {\bibinfo {author} {\bibfnamefont {M.~J.~D.}\
  \bibnamefont {Powell}},\ }\bibfield  {title} {\bibinfo {title} {On fast trust
  region methods for quadratic models with linear constraints},\ }\href
  {https://doi.org/10.1007/s12532-015-0084-4} {\bibfield  {journal} {\bibinfo
  {journal} {Mathematical Programming Computation}\ }\textbf {\bibinfo {volume}
  {7}},\ \bibinfo {pages} {237} (\bibinfo {year} {2015})}\BibitemShut {NoStop}%
\bibitem [{\citenamefont {Huyer}\ and\ \citenamefont
  {Neumaier}(2008)}]{SNOBFIT}%
  \BibitemOpen
  \bibfield  {author} {\bibinfo {author} {\bibfnamefont {W.}~\bibnamefont
  {Huyer}}\ and\ \bibinfo {author} {\bibfnamefont {A.}~\bibnamefont
  {Neumaier}},\ }\bibfield  {title} {\bibinfo {title} {{SNOBFIT--Stable} noisy
  optimization by branch and fit},\ }\href
  {https://doi.org/10.1145/1377612.1377613} {\bibfield  {journal} {\bibinfo
  {journal} {ACM Transactions on Mathematical Software}\ }\textbf {\bibinfo
  {volume} {35}},\ \bibinfo {pages} {1} (\bibinfo {year} {2008})}\BibitemShut
  {NoStop}%
\bibitem [{\citenamefont {DeCross}\ \emph {et~al.}(2023)\citenamefont
  {DeCross}, \citenamefont {Chertkov}, \citenamefont {Kohagen},\ and\
  \citenamefont {Foss-Feig}}]{decross2023qubit}%
  \BibitemOpen
  \bibfield  {author} {\bibinfo {author} {\bibfnamefont {M.}~\bibnamefont
  {DeCross}}, \bibinfo {author} {\bibfnamefont {E.}~\bibnamefont {Chertkov}},
  \bibinfo {author} {\bibfnamefont {M.}~\bibnamefont {Kohagen}},\ and\ \bibinfo
  {author} {\bibfnamefont {M.}~\bibnamefont {Foss-Feig}},\ }\bibfield  {title}
  {\bibinfo {title} {Qubit-reuse compilation with mid-circuit measurement and
  reset},\ }\href {https://doi.org/10.1103/PhysRevX.13.041057} {\bibfield
  {journal} {\bibinfo  {journal} {Physical Review X}\ }\textbf {\bibinfo
  {volume} {13}},\ \bibinfo {pages} {041057} (\bibinfo {year}
  {2023})}\BibitemShut {NoStop}%
\bibitem [{\citenamefont {Harrigan}\ \emph {et~al.}(2021)\citenamefont
  {Harrigan}, \citenamefont {Sung}, \citenamefont {Neeley}, \citenamefont
  {Satzinger}, \citenamefont {Arute}, \citenamefont {Arya}, \citenamefont
  {Atalaya}, \citenamefont {Bardin}, \citenamefont {Barends}, \citenamefont
  {Boixo} \emph {et~al.}}]{harrigan2021quantum}%
  \BibitemOpen
  \bibfield  {author} {\bibinfo {author} {\bibfnamefont {M.~P.}\ \bibnamefont
  {Harrigan}}, \bibinfo {author} {\bibfnamefont {K.~J.}\ \bibnamefont {Sung}},
  \bibinfo {author} {\bibfnamefont {M.}~\bibnamefont {Neeley}}, \bibinfo
  {author} {\bibfnamefont {K.~J.}\ \bibnamefont {Satzinger}}, \bibinfo {author}
  {\bibfnamefont {F.}~\bibnamefont {Arute}}, \bibinfo {author} {\bibfnamefont
  {K.}~\bibnamefont {Arya}}, \bibinfo {author} {\bibfnamefont {J.}~\bibnamefont
  {Atalaya}}, \bibinfo {author} {\bibfnamefont {J.~C.}\ \bibnamefont {Bardin}},
  \bibinfo {author} {\bibfnamefont {R.}~\bibnamefont {Barends}}, \bibinfo
  {author} {\bibfnamefont {S.}~\bibnamefont {Boixo}}, \emph {et~al.},\
  }\bibfield  {title} {\bibinfo {title} {Quantum approximate optimization of
  non-planar graph problems on a planar superconducting processor},\ }\href
  {https://doi.org/10.1038/s41567-020-01105-y} {\bibfield  {journal} {\bibinfo
  {journal} {Nature Physics}\ }\textbf {\bibinfo {volume} {17}},\ \bibinfo
  {pages} {332} (\bibinfo {year} {2021})}\BibitemShut {NoStop}%
\bibitem [{H1()}]{H1}%
  \BibitemOpen
  \href@noop {} {\bibinfo {title} {Quantinuum {H1-1}}},\ \bibinfo
  {howpublished} {\url{https://www.quantinuum.com/}},\ \bibinfo {note} {{Nov.}
  10 - {Nov.} 28, 2023}\BibitemShut {NoStop}%
\bibitem [{H2()}]{H2}%
  \BibitemOpen
  \href@noop {} {\bibinfo {title} {Quantinuum {H2-1}}},\ \bibinfo
  {howpublished} {\url{https://www.quantinuum.com/}},\ \bibinfo {note} {{Jan.}
  10 - {Jan.} 23, 2024}\BibitemShut {NoStop}%
\bibitem [{\citenamefont {Lykov}\ \emph {et~al.}(2023)\citenamefont {Lykov},
  \citenamefont {Shaydulin}, \citenamefont {Sun}, \citenamefont {Alexeev},\
  and\ \citenamefont {Pistoia}}]{Lykov2023}%
  \BibitemOpen
  \bibfield  {author} {\bibinfo {author} {\bibfnamefont {D.}~\bibnamefont
  {Lykov}}, \bibinfo {author} {\bibfnamefont {R.}~\bibnamefont {Shaydulin}},
  \bibinfo {author} {\bibfnamefont {Y.}~\bibnamefont {Sun}}, \bibinfo {author}
  {\bibfnamefont {Y.}~\bibnamefont {Alexeev}},\ and\ \bibinfo {author}
  {\bibfnamefont {M.}~\bibnamefont {Pistoia}},\ }\bibfield  {title} {\bibinfo
  {title} {Fast simulation of high-depth {QAOA} circuits},\ }in\ \href
  {https://doi.org/10.1145/3624062.3624216} {\emph {\bibinfo {booktitle}
  {Proceedings of the SC '23 Workshops of The International Conference on High
  Performance Computing, Network, Storage, and Analysis}}},\ \bibinfo {series
  and number} {SC-W 2023}\ (\bibinfo  {publisher} {ACM},\ \bibinfo {year}
  {2023})\BibitemShut {NoStop}%
\bibitem [{\citenamefont {Hao}\ \emph {et~al.}(2023)\citenamefont {Hao},
  \citenamefont {Liu},\ and\ \citenamefont {Tannu}}]{OSCAR}%
  \BibitemOpen
  \bibfield  {author} {\bibinfo {author} {\bibfnamefont {T.}~\bibnamefont
  {Hao}}, \bibinfo {author} {\bibfnamefont {K.}~\bibnamefont {Liu}},\ and\
  \bibinfo {author} {\bibfnamefont {S.}~\bibnamefont {Tannu}},\ }\bibfield
  {title} {\bibinfo {title} {Enabling high performance debugging for
  variational quantum algorithms using compressed sensing},\ }in\ \href
  {https://doi.org/10.1145/3579371.3589044} {\emph {\bibinfo {booktitle}
  {Proceedings of the 50th Annual International Symposium on Computer
  Architecture}}}\ (\bibinfo  {publisher} {Association for Computing
  Machinery},\ \bibinfo {address} {New York, NY, USA},\ \bibinfo {year}
  {2023})\BibitemShut {NoStop}%
\bibitem [{dat()}]{datalink}%
  \BibitemOpen
  \href@noop {} {}\bibinfo {note}
  {\url{https://doi.org/10.5281/zenodo.12209739}}\BibitemShut {NoStop}%
\bibitem [{cod()}]{codelink}%
  \BibitemOpen
  \href@noop {} {}\bibinfo {note}
  {\url{https://github.com/jpmorganchase/End-to-End_Protocol_for_High-Quality_QAOA_Parameters}}\BibitemShut
  {NoStop}%
\end{thebibliography}%

\section*{Disclaimer}
This paper was prepared for informational purposes with contributions from the Global Technology Applied Research center of JPMorganChase. This paper is not a product of the Research Department of JPMorganChase or its affiliates. Neither JPMorganChase nor any of its affiliates makes any explicit or implied representation or warranty and none of them accept any liability in connection with this position paper, including, without limitation, with respect to the completeness, accuracy, or reliability of the information contained herein and the potential legal, compliance, tax, or accounting effects thereof. This document is not intended as investment research or investment advice, or as a recommendation, offer, or solicitation for the purchase or sale of any security, financial instrument, financial product or service, or to be used in any way for evaluating the merits of participating in any transaction.

The submitted manuscript includes contributions from UChicago Argonne, LLC, Operator of Argonne National Laboratory (``Argonne''). Argonne, a U.S.\ Department of Energy Office of Science laboratory, is operated under Contract No.\ DE-AC02-06CH11357. The U.S.\ Government retains for itself, and others acting on its behalf, a paid-up nonexclusive, irrevocable worldwide license in said article to reproduce, prepare derivative works, distribute copies to the public, and perform publicly and display publicly, by or on behalf of the Government.  The Department of Energy will provide public access to these results of federally sponsored research in accordance with the DOE Public Access Plan \url{http://energy.gov/downloads/doe-public-access-plan}.

\onecolumngrid
\appendix
\section{Additional results}\label{sec:appendix}
\begin{figure*}[h]
    \centering
    \begin{subfigure}{0.245\textwidth}
        \centering
        \includegraphics[width=\textwidth]{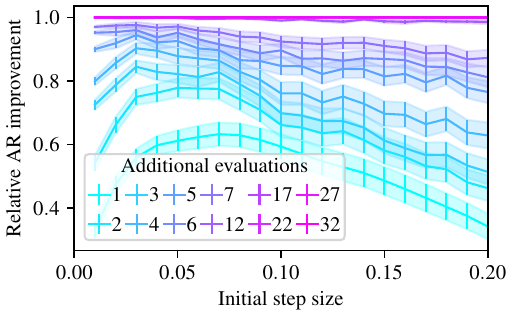}
        \subcaption{$p=1$ MaxCut}
    \end{subfigure}
    \begin{subfigure}{0.245\textwidth}
        \centering
        \includegraphics[width=\textwidth]{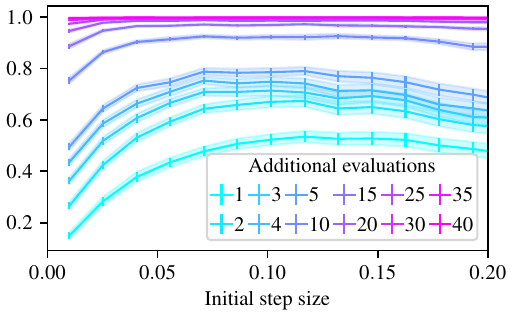}
        \subcaption{$p=2$ MaxCut}
    \end{subfigure}
    \begin{subfigure}{0.245\textwidth}
        \centering
        \includegraphics[width=\textwidth]{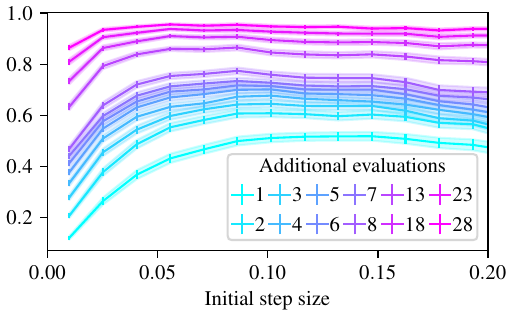}
        \subcaption{$p=3$ MaxCut}
    \end{subfigure}
    \begin{subfigure}{0.245\textwidth}
        \centering
        \includegraphics[width=\textwidth]{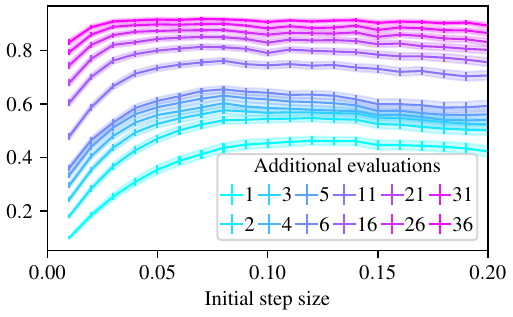}
        \subcaption{$p=4$ MaxCut}
    \end{subfigure}
    \begin{subfigure}{0.245\textwidth}
        \centering
        \includegraphics[width=\textwidth]{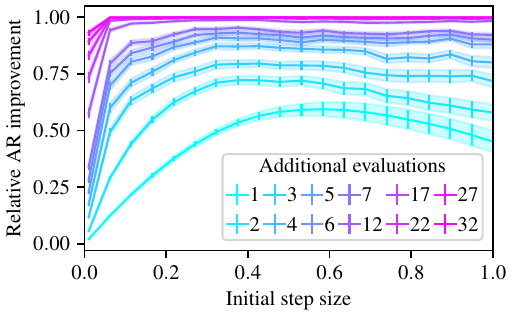}
        \subcaption{$p=1$ PO}
    \end{subfigure}
    \begin{subfigure}{0.245\textwidth}
        \centering
        \includegraphics[width=\textwidth]{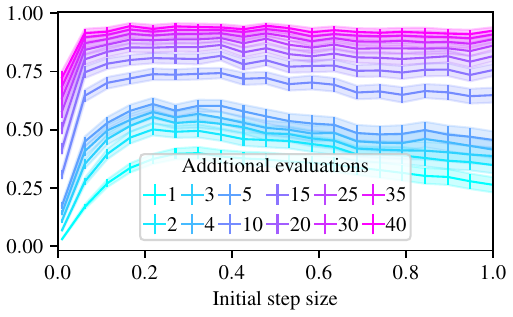}
        \subcaption{$p=2$ PO}
    \end{subfigure}
    \begin{subfigure}{0.245\textwidth}
        \centering
        \includegraphics[width=\textwidth]{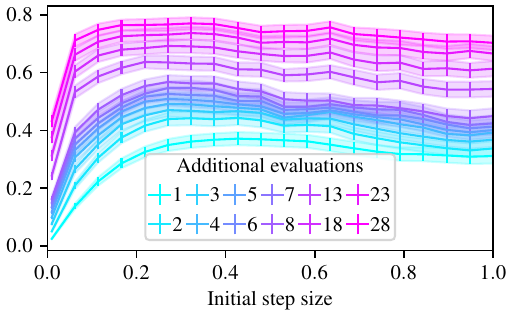}
        \subcaption{$p=3$ PO}
    \end{subfigure}
    \begin{subfigure}{0.245\textwidth}
        \centering
        \includegraphics[width=\textwidth]{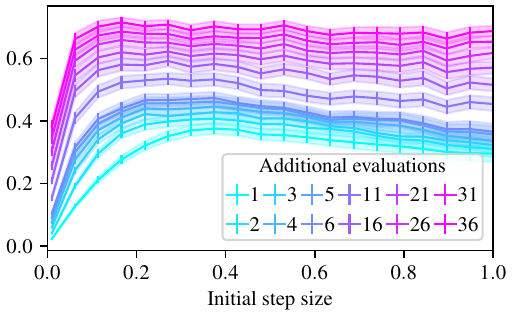}
        \subcaption{$p=4$ PO}
    \end{subfigure}
    \caption{Mean relative AR improvement (with standard error over instances) of COBYLA on $p\in \{1,2,3,4\}$ MaxCut and PO instances as a function of initial step size. The label of each line represents the number of function evaluations allowed after the initial evaluations. We observe that with a given problem and initialization strategy, COBYLA is generally not sensitive to the initial step size or QAOA depth.} 
    \label{fig:more_rhobeg}
\end{figure*}

\begin{figure*}[h]
    \centering
    \begin{subfigure}{0.245\textwidth}
        \centering
        \includegraphics[width=\textwidth]{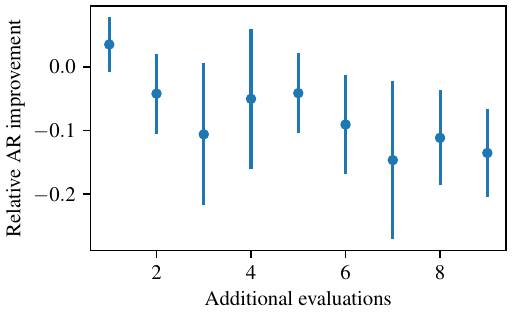}
        \subcaption{COBYLA $p=2$}
    \end{subfigure}
    \begin{subfigure}{0.245\textwidth}
        \centering
        \includegraphics[width=\textwidth]{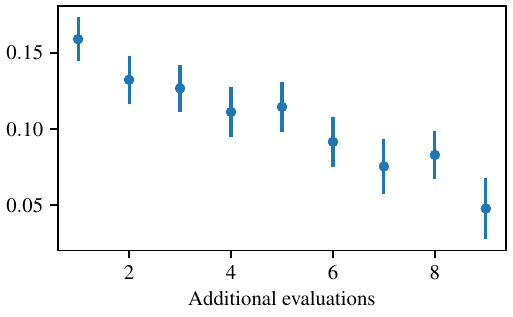}
        \subcaption{COBYLA $p=3$}
    \end{subfigure}
    \begin{subfigure}{0.245\textwidth}
        \centering
        \includegraphics[width=\textwidth]{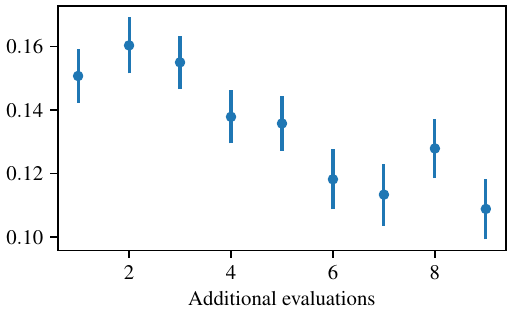}
        \subcaption{COBYLA $p=4$}
    \end{subfigure}
    \begin{subfigure}{0.245\textwidth}
        \centering
        \includegraphics[width=\textwidth]{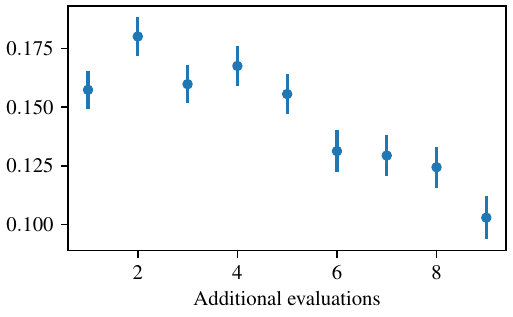}
        \subcaption{COBYLA $p=5$}
    \end{subfigure}
    \begin{subfigure}{0.245\textwidth}
        \centering
        \includegraphics[width=\textwidth]{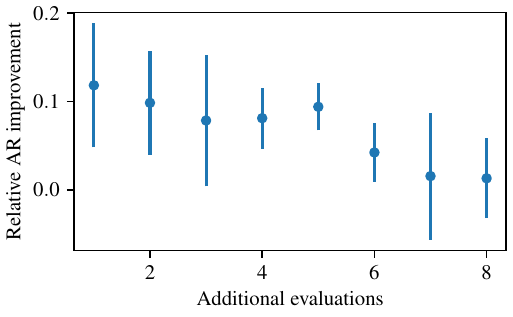}
        \subcaption{Py-BOBYQA $p=2$}
    \end{subfigure}
    \begin{subfigure}{0.245\textwidth}
        \centering
        \includegraphics[width=\textwidth]{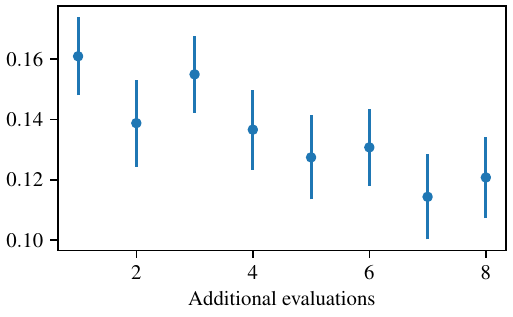}
        \subcaption{Py-BOBYQA $p=3$}
    \end{subfigure}
    \begin{subfigure}{0.245\textwidth}
        \centering
        \includegraphics[width=\textwidth]{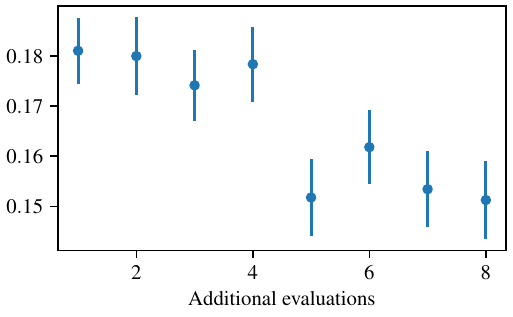}
        \subcaption{Py-BOBYQA $p=4$}
    \end{subfigure}
    \begin{subfigure}{0.245\textwidth}
        \centering
        \includegraphics[width=\textwidth]{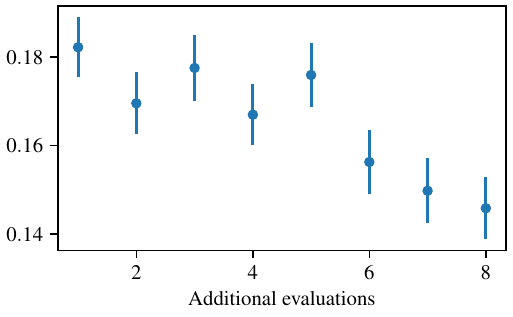}
        \subcaption{Py-BOBYQA $p=5$}
    \end{subfigure}
    \caption{Mean relative AR improvement (with standard error over instances) of optimizing 1,000 $p\in \{2,3,4,5\}\ n=12$ MaxCut instances as a function of the number of additional evaluations after the first $2p+1$/$4p+1$ initial evaluations for COBYLA/Py-BOBYQA. 
    }
    \label{fig:more_budget_slice}
\end{figure*}

\begin{figure*}[h]
    \centering
    \begin{subfigure}{0.245\textwidth}
        \centering
        \includegraphics[width=\textwidth]{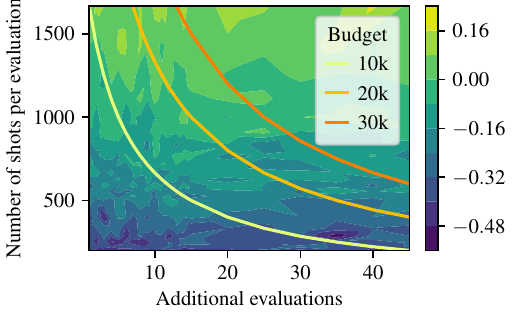}
        \subcaption{COBYLA $p=2$}
    \end{subfigure}
    \begin{subfigure}{0.245\textwidth}
        \centering
        \includegraphics[width=\textwidth]{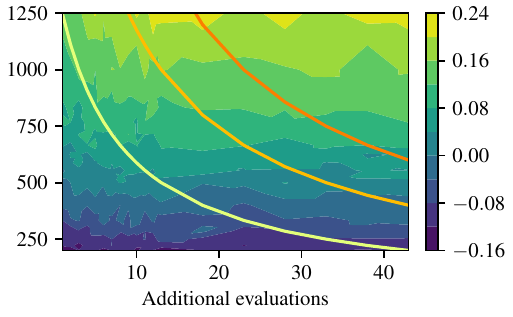}
        \subcaption{COBYLA $p=3$}
    \end{subfigure}
    \begin{subfigure}{0.245\textwidth}
        \centering
        \includegraphics[width=\textwidth]{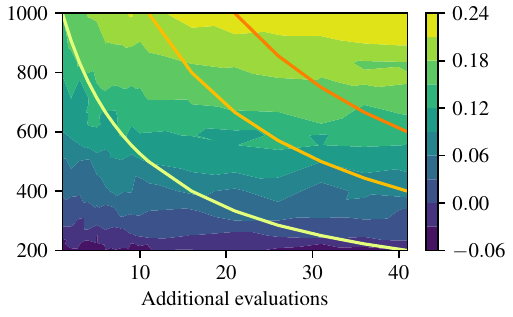}
        \subcaption{COBYLA $p=4$}
    \end{subfigure}
    \begin{subfigure}{0.245\textwidth}
        \centering
        \includegraphics[width=\textwidth]{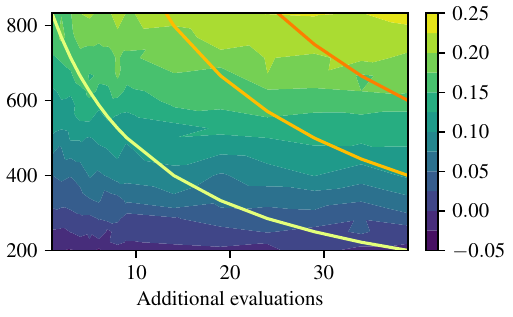}
        \subcaption{COBYLA $p=5$}
    \end{subfigure}
    \begin{subfigure}{0.245\textwidth}
        \centering
        \includegraphics[width=\textwidth]{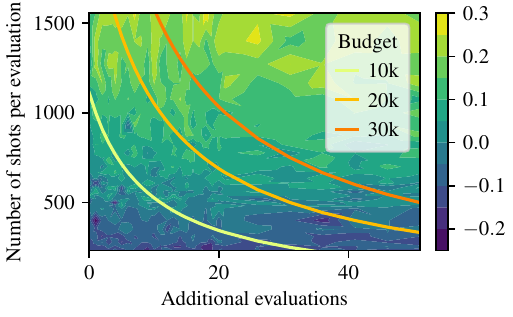}
        \subcaption{Py-BOBYQA $p=2$}
    \end{subfigure}
    \begin{subfigure}{0.245\textwidth}
        \centering
        \includegraphics[width=\textwidth]{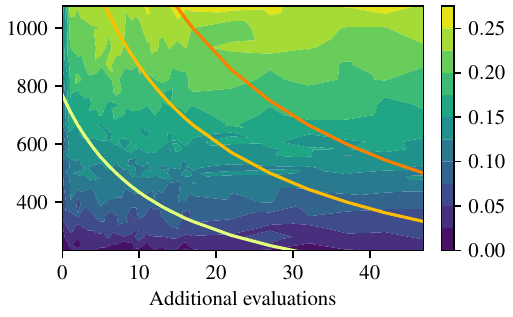}
        \subcaption{Py-BOBYQA $p=3$}
    \end{subfigure}
    \begin{subfigure}{0.245\textwidth}
        \centering
        \includegraphics[width=\textwidth]{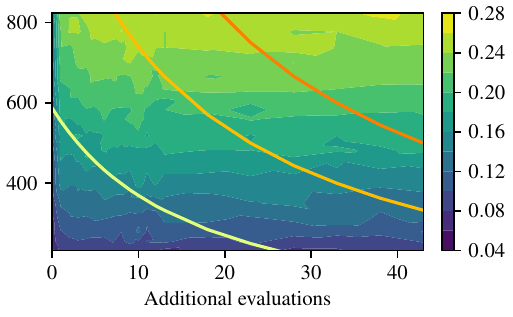}
        \subcaption{Py-BOBYQA $p=4$}
    \end{subfigure}
    \begin{subfigure}{0.245\textwidth}
        \centering
        \includegraphics[width=\textwidth]{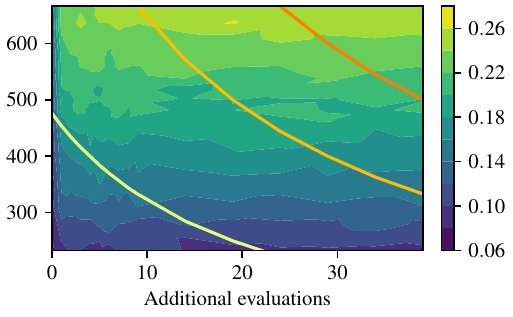}
        \subcaption{Py-BOBYQA $p=5$}
    \end{subfigure}
    \caption{Contour plot of optimizing 1,000 $p\in \{2,3,4,5\}\ n=12$ MaxCut instances spanned by the number of additional evaluations and the number of shots per evaluation. The color represents mean relative AR improvement, and the three lines correspond to a total budget of 10k, 20k, and 30k, respectively.} 
    \label{fig:more_budget_heatmap}
\end{figure*}

Here, we show additional results we have obtained, including the initial step size study for $p\in \{1,2,3,4\}$ (\Cref{fig:more_rhobeg}) and the budget allocation study for $p\in \{2,3,4,5\}$ (\Cref{fig:more_budget_slice} and \Cref{fig:more_budget_heatmap}).

In \Cref{fig:more_budget_slice}, following the similar setup as \Cref{fig:budget-slice}, we did an additional comparison between COBYLA implemented in NLopt~\cite{NLopt} and Py-BOBYQA~\cite{cartis2019improving}.
The contours of their performance under different budgets are shown in \Cref{fig:more_budget_heatmap}. 
Py-BOBYQA is an improved version of BOBYQA implemented in Python. 
In our numerical experiments, we found that the performance of Py-BOBYQA is similar to the standard BOBYQA implemented in NLopt~\cite{NLopt}.
We did not enable the \texttt{objfun\_has\_noise} flag in Py-BOBYQA due to its effect of defaulting to $\frac{1}{2}(2p+1)(2p+2)$ initial function evaluations (instead of $4p+1$) and using multi-restarts, both of which contradicted our shot-frugal setting. We also wanted to be consistent with the $4p+1$ initial function evaluations we used for BOBYQA in our optimizer comparison experiment (\Cref{fig:optimizer}).

Note that for $p=1$, the initial points are of very high quality, and the maximum achievable approximation ratio (AR) is relatively low due to the shallow QAOA depth. Consequently, the improvable AR is very small, and the optimizer struggles to improve beyond the quality of the initial point, especially with a highly stochastic objective resulting from the low shot budget. Therefore, we do not show $p=1$ figures or use $p=1$ MaxCut in the optimizer benchmarking experiments. 

For a small $p$, the difference in the number of initial evaluations does not result in a huge disparity in the number of shots per evaluation. At $p=2$, Py-BOBYQA outperforms COBYLA with its quadratic model. For $p\in \{3,4,5\}$, their performances are comparable, and COBYLA shows a progressive momentum.

\end{document}